\documentclass{article}
\usepackage{amsmath}
\usepackage{graphicx}
\usepackage{psfrag}
\usepackage{setspace}
\newcommand{\ka}{\kappa}

\renewcommand{\vec}[1]{\mathbf{#1}}
\newcommand{\mc}[1]{\mathcal{#1}}

\newcommand{\la}{\Lambda}
\newcommand{\om}{\Omega}
\newcommand{\pa}{\partial}

\renewcommand{\dag}{\dagger}
\def\Ib5{{\hbox{\rlap{\hbox{\raise.27ex\hbox{-}}}I}}^{(5)}}
\def\ba{\bar{a}}
\def\bb{\bar{b}}
\def\be{\beta}
\def\al{\alpha}
\renewcommand{\b}[1]{\bar{#1}}
\def\Ib{{\hbox{\rlap{\hbox{\raise.27ex\hbox{-}}}I}}}
\def\F{\mathcal{F}}
\def\W{\mathcal{W}}
\def\M{\mathcal{M}}
\def\mI{\mathcal{I}}

\newcommand{\ga}{\gamma}
\newcommand{\tp}{\tilde{p}}
\newcommand{\ta}{\tilde{a}}
\newcommand{\tr}{\tilde{r}}
\newcommand{\tR}{\tilde{R}}
\newcommand{\eps}{\epsilon}

\begin{document}

\title{Dynamical evolution and leading order gravitational wave
emission of Riemann-S binaries}
\author{D\"orte Hansen\footnote{e-mail: nch@tpi.uni-jena.de}\\
        Theoretisch-Physikalisches Institut \\
        Friedrich-Schiller-Universit\"at Jena\\
        Max-Wien-Platz 1\\
        07743 Jena\\
        Germany
        }
\date{}
\maketitle

%\doublespacing

\begin{abstract}
An approximate strategy for studying the evolution of binary systems
of extended objects is introduced. The stars are assumed to be
polytropic ellipsoids. The surfaces of constant density maintain
their ellipsoidal shape during the time evolution. The equations of
hydrodynamics then reduce to a system of ordinary differential
equations for the internal velocities, the principal axes of the stars
and the orbital parameters. The equations of motion are given within
Lagrangian and Hamiltonian formalism. The special case when both stars
are axially symmetric fluid configurations is considered. Leading order
gravitational radiation reaction is incorporated, where the
quasi-static approximation is applied to the internal degrees of
freedom of the stars. The influence of the stellar parameters,
in particular the influence of the polytropic index $n$, on
the leading order gravitational waveforms is studied.
\end{abstract}


\section{Introduction}

The search for gravitational waves is one of the most challenging projects
of 21st century physics. Inspiraling compact binaries are among the
most promising sources for gravitational waves that could be detected
by gravitational waves observatories, such as LIGO, VIRGO and
GEO600. During most of the inspiral time the gravitational waves
emitted by a compact binary system are much to weak, the frequencies
are much to small to be detectable by existing gravitational wave
observatories. However, at orbital distances corresponding to  the
last 10 or 15 minutes of the inspiral time, the orbital frequency
increases from around 1 Hz up to 1000 Hz, while in the same time
the amplitude of the gravitational waves increases, thus improving
the chance of actually detecting the weak signals.  \\
In order to extract a possible signal from the  detector noise the
gravitational waveforms emitted by inspiraling binary systems must be
known in great detail, in particular during the last stages of the
inspiral process before the final plunge.
The frequency gap covered by contemporary gravitational wave
observatories ranges from 1 to 1000 Hz. In this regime post-Newtonian (pN)
effects become important and have to be included into the
analysis. For non spinning point-particle binaries an analytic solution
is available up to the 3rd pN approximation 
\cite{Memmesheimer:2004cv}. Incorporating the spin leads to enormeous
complifications. Until now there exists a solution only for point-mass
binaries with either two equal masses and arbitrary spins, or two
arbitrary masses and only one spinning object \cite{Konigsdorffer:2005sc}.\\
However, while the components of black hole-black hole
binaries can be treated as pointlike objects eventually up to the innermost
stable circular orbit (ISCO),
yet another effect has to be incorporated into the analysis for all
other binary systems: the internal structure of the components.
 Lai and Wiseman argue that during most of the final
inspiral process even the neutron stars can be treated as pointlike
objects \cite{Lai:1996sv}. This is true if one considers only a few orbits. If,
however, one is interested in the long-term evolution of the system,
the tidal interaction of  the neutron stars will lead to a phase
shift in the gravitational waveforms, which is not neglegible anymore.\\
In this paper we shall focus on the problem of
including tidal interaction into the equations of motion and we
discuss the influence of the internal stellar dynamics on the
binary's dynamics as well as on the  gravitational waveforms in
leading order approximation. Post-Newtonian effects will be included
in a forthcoming paper.\\
A detailed description of the tidal interaction and the corresponding
changes in the gravitational waveforms requires the application of
 three-dimensional numerical hydrodynamics. This is well beyond the
 scope of this paper. Instead we shall consider an approximative formulation of
the problem. This strategy allows an analytic formulation of the
equations of motion. Moreover,  we are able to  run long-term
calculations of the system's dynamics with sufficient accuracy.
There are basically two approaches for an approximate description of
the tidal interaction. The first one is based on a linear adiabatic
theory, describing small derivations from equilibrium. Assuming an
adiabatic equation of state, the linearized
equations of hydrodynamics have been derived by Ledoux and Walraven
\cite{Ledoux}
and Dziembowski \cite{Dziembowski:1971}. Press and Teukolsky
\cite{Press:1977} developed a technique for calculating normal modes
of non-radial stellar oscillations in the presence of tidal forces.
Using the formulation of Dziembowski, Kokkotas and Sch\"afer
studied the leading order dynamics and gravitational waveforms of an NS-NS
binary system, in particular taking into account the influence of
tidal interaction on the actual waveforms
\cite{Kokkotas:1995xe}. Resonant tidal excitations of a binary neutron
star with polytropic equations of state have been investigated by Ho
in leading order approximation \cite{Ho:1998hq}.    \\
However, there are  scenarios when the linear adiabatic theory is not
applicable. In particular the theory fails when the stellar
oscillations cannot be considered as small derivations from
equilibrium. Thus far only little investigation has been undertaken in
this direction. For a rotating, oscillating disk of dust there exists an
analytic solution up to first post-Newtonian order
\cite{Schafer:1994a}. A system of an rotating, osillating
dusty disk and a point-mass object has been studied by the author in
a recent paper \cite{Hansen:2005}. A so-called 'affine stellar model' for
polytropic, ellipsoidal configurations has been developed by
Carter and Luminet \cite{Carter:1983}. Here the density contours in
the star are assumed to form homologeous ellipsoids. The model allows
for the discussion of arbitrary large amplitudes of the tidally
generated oscillations, but fails to give accurate results for small
oscillations. The excited mode roughly corresponds to the $f$-mode of
the oscillations (see also \cite{Kosovichev:1992}).  Later
on Lai and Shapiro extended this model to allow for the investigation
of Riemann-S binaries \cite{Lai:1995a}. In this paper we shall
follow the approach of Lai and Shapiro, pointing out some
discrepancies in the authors' formulation
of the equations of motion as well as for the gravitational waveforms.
\\
The paper is organized as follows: 
In section 2 we briefly recapitulate the Lagrangian formulation of
the dynamics of Riemann-S binaries derived by Lai and Shapiro in a
series of papers \cite{Lai:1995a}, \cite{Lai:1994hf},
\cite{Lai:1993pa}. In this model the stars are considered as rotating
and oscillating triaxial ellipsoids with a polytropic equation of
state. To keep the analysis as simple as possible, the tidal
interaction potential is truncated, only the leading order
(i.e. quadrupole) tidal interaction is taken into account. For later
reasons we shall also present the
Hamiltonian formulation of this problem. In particular we shall
specialize to axially symmetric polytropic stars\footnote{Just setting
$a_1=a_2$ in the equations of motions gives rise to divergent terms.}.
Section 3 is devoted to the derivation of the leading order
gravitational reaction terms appearing in the dynamical equations. We
shall point out some discrepancies between our approach and the
approach taken by Lai and Shapiro. The leading order gravitational
waveforms are derived in section 4. In particular we shall apply the
quasi-static approximation to the internal stellar dynamics. The
corresponding equations are derived for general triaxial Riemann-S binaries,
but for simplicity we restrict ourselves to axially symmetric
configurations in the numerical calculations. The numerical results
and some possible physical applications of our model are discussed in
section 5.  In particular we study the influence of the polytropic
index and hence of the equation of state on the dynamics as well as on
the leading order gravitational waveforms. We shall show that the
effect of the internal stellar structure on the long term
evolution of the binary is quite significant. We argue that the
exact knowledge of the polytropic equation of state is essential in
order to calculate
the leading order gravitational waveforms emitted by the binary system
correctly.

\section{The Newtonian dynamics  of Riemann-S binaries}

\subsection{Lagrangian formalism}

On Newtonian order the Lagrangian equations of motion for compressible
Riemann-S binaries have been derived by Lai and Shapiro in several
papers (see \cite{Lai:1995a}, \cite{Lai:1994hf},
\cite{Lai:1993pa}) and we shall therefore only shortly recapitulate their
strategy. We then turn our attention to the Hamiltonian formulation of
the problem which will be more suitable for numerical
calculations. The case of axially symmetric, polytropic stars is of
particular interest for us, but requires some special care in order not
to overcount the degrees of freedom.\\
Consider a binary consisting of two ellipsoidal fluid configurations
$M$ and $M'$, asssuming the stellar matter to obey a polytropic equation of
state,
\begin{align}
\label{poly}
     P=K\,\rho^{1+1/n},\qquad
     P'=K'\rho'^{1+1/n'}.
\end{align}
Here $n$ and $n'$ are the polytropic indices, while $K$ and $K'$ represent
constants which are determined by the equilibrium radii of
non-rotating, spherically symmetric polytropes of the same mass. A
Riemann-S ellipsoid is characterized by the angular velocity
$\vec\om=\om\vec e_z$ of the ellipsoidal figure around the principal
axes and the internal vorticity $\mathbf{\zeta=}\zeta\vec e_z$. The
model is based on two assumptions: First, we shall require
the vorticity to be uniform, and second, the surfaces of
constant density inside the stars are assumed to form self-similar
ellipsoids. In other words, each star is described by only five
degrees of freedom: the three semi-major axes $a_1,a_2,a_3$ and
$a'_1,a'_2,a'_3$ respectively, and the two angles $\psi$ and $\ga$,
which are introduced in figure 1.\\
Consider now an isolated star $M$. If the assumptions given above are
to be fullfilled, it is immediately clear that the fluid velocity in the body
fixed system, which rotates with angular velocity $\vec\om=\om \vec
e_3$ ($\om=\dot{\ga}$) with respect to an inertial system centered
at $M$,  obeys the following ansatz:
\begin{align}
\label{intern1}
      \vec u_c=Q_1x_2\vec e_1+Q_2x_1\vec e_2.
\end{align}
Here $Q_1,Q_2$ are constants and $x_1,x_2$ denote coordinates in the
body fixed frame. The second requirement yields
\begin{align*}
     Q_1=-\frac{a_1^2}{a_1^2+a_2^2}\,\zeta=\frac{a_1}{a_2}\,\la,\qquad
     Q_2=\frac{a_2^2}{a_1^2+a_2^2}\,\zeta=-\frac{a_2}{a_1}\,\la,
\end{align*}
where $\zeta$ is the vorticity in the body fixed frame and
$\la=\dot{\psi}$ is the angular velocity of the internal fluidal
motion. On the other hand, observing the stellar matter in an inertial
system centered at $M$ it's velocity is given by 
\begin{align}
\label{intern2}
     \vec u_{IS}=\vec u_c+\vec\om\wedge\vec r.
\end{align} 
The rotational energy of the star reads
\begin{align}
\label{intern3}
     T_{rot}&=\frac{1}{2}\int \rho\,\vec u_{IS}\cdot\vec u_{IS}dV
     \nonumber \\
     &=\frac{\ka_nM}{10}(a_1^2+a_2^2)(\la^2+\om^2)-\frac{2}{5}\ka_nM a_1a_2\la\om,
\end{align}
where  $\ka_n\le 1$ is a constant, which can be obtained from the
Lane-Emden equation  (see Appendix).
It can be easily seen that the inertial tensor $I_{ij}=\int \rho
x_ix_j dV$ in the body fixed system takes a very simple form,
\begin{align}
\label{inertialtens1}
     I_{ij}=\int\rho x_ix_jdV=\frac{\ka_nM}{5}a_i^2\delta_{ij},\qquad
      \ka_n:=\frac{5}{3}\frac{\int_0^{\xi_1} \theta^n \xi^4 d\xi}{\xi_1^4|\theta'_1|}.
\end{align}
In particular, for $a_1\equiv a_2$ the rotational energy reduces to
\begin{align}
      T_{rot}=\frac{\ka_nM}{5}(\om-\la)^2a_1^2.
\end{align}
The total kinetic energy of star $M$ is  given by
\begin{align}
\label{kin}
    T_s=\frac{\ka_nM}{10}(a_1^2+a_2^2)(\om^2+\la^2)-\frac{2}{5}\ka_nMa_1a_2\la\om +\frac{\ka_nM}{10}(\dot{a}_1^2+\dot{a}_2^2+\dot{a}_3^2),
\end{align}
and the Lagrangian $L_s$ of the star reads
\begin{align}
\label{lagstar1}
   L_s=T_s-U-W.
\end{align}
The gravitational self-energy $W$, and $U$, the internal energy
of the star, have been computed by \cite{Lai:1995a}. They are given by
\begin{align}
     U=\int n\,\frac{P}{\rho}\,dm=k_1KM\rho_c^{1/n},
\end{align}
where $k_1=\frac{n(n+1)}{5-n}\xi_1|\theta'_1|$ is constant, and
\begin{align}
     W=-\frac{3}{5-n}\frac{GM^2}{2R^3}\mc J.
\end{align}
Here $R=(a_1a_2a_3)^{1/3}$ is the mean radius of the ellipsoid and
$\mc J$ is given by $ \mc J=a_1^2A_1+a_2^2A_2+a_3^2A_3$. The
coefficients $A_i$, defined by Chandrasekhar
\cite{Chandrasekhar:1969}, are given in the Appendix.\\
Now let us come back to the binary system. The tidal interaction
potential is clearly dominated by the quadrupole interaction, which
will be the only terms to be included into our calculations. Thus the
orbital Lagrangian reads
\begin{align}
    L_{orb}=\frac{\mu}{2}\dot{r}^2+\frac{\mu r^2}{2}\dot{\phi}^2-W_{int},
\end{align}
where $\mu=MM'/(M+M')$ is the reduced mass, and
the interaction potential $W_{int}$ is
\begin{align}
\label{Wint}
     W_{int}&=-\frac{GMM'}{r}-\frac{GMM'\ka_n}{10r^3}\bigl[
     a_1^2(3\cos^2\al-1)+a_2^2(3\sin^2\al-1)-a_3^2\bigr] \nonumber \\
     &\hspace{0.5cm}
     -\frac{GMM'\ka_n'}{10r^3}\bigl[a_1'^2(3\cos^2\al'-1)+a_2'^2(3\sin^2\al'-1)-a_3'^2\bigr].    
\end{align}
The meaning of the angle $\al=\phi-\ga$ can be read off from Fig.
1. It is now easy to write down the Lagrangian of a general Riemann-S
binary according to
\begin{align}
\label{lag1}
     L&=\frac{\mu}{2}\dot{r}^2+\frac{\mu
     r^2}{2}\dot{\phi}^2+\frac{GMM'}{r}+\frac{GMM'\ka_n}{10r^3}\bigl[
     a_1^2(3\cos^2\al-1)+a_2^2(3\sin^2\al-1)-a_3^2)\bigr] \nonumber \\
     &\hspace{4.2cm}
     +\frac{GMM'\ka_n'}{10r^3}\bigl[a_1'^2(3\cos^2\al'-1)+a_2'^2(3\sin^2\al'-1)-a_3'^2\bigr] \nonumber \\
     &\hspace{0.5cm}
     +\frac{\ka_nM}{10}(a_1^2+a_2^2)(\la^2+\om^2)-\frac{2}{5}\ka_nMa_1a_2\la\om +\frac{\ka_nM}{10}(\dot{a}_1^2+\dot{a}_2^2+\dot{a}_3^2) -k_1KM\rho_c^{1/n}
     \nonumber \\
     &\hspace{0.5cm}+\frac{3}{5-n}\frac{GM^2}{2R^3}\mc J
     +\frac{\ka_n'M'}{10}(a_1'^2+a_2'^2)(\om'^2+\la'^2)-\frac{2}{5}\ka_n'M'a_1'a_2'\la'\om'\nonumber \\
     &\hspace{0.5cm}
     +\frac{\ka_n'M'}{10}(\dot{a}_1'^2+\dot{a}_2'^2+\dot{a}_3'^2)-k_1'K'M'\rho_c'^{1/n'}+\frac{3}{5-n'}\frac{GM'^2}{2R'^3}\mc J'.
\end{align}
It is  straightforward to derive the equations of motion governing
the dynamics of the binary system. Let us first focus on the most
general case when all three semi-major axes $a_i$ are
different. With the central density $\rho_c$ being proportional to
$1/(a_1a_2a_3)$ and using
\begin{align*}
    \frac{\pa \mc J}{\pa a_i}=\frac{1}{a_i}(\mc J-a_i^2A_i)
\end{align*}   
(see Appendix) the equations of motion derived from the Lagrangian
(\ref{lag1}) read
\begin{align}
\label{lageom11}
     \ddot{a}_1&=\frac{GM'}{r^3}\,a_1(3\cos^2\al-1)+a_1(\om^2+\la^2)-2a_2\la\om+\left(\frac{5k_1}{n\ka_n}\frac{P_c}{\rho_c}\right)\frac{1}{a_1}
     -\frac{3GM}{\ka_n(1-\frac{n}{5})}\frac{a_1A_1}{2R^3} ,\\
\label{lageom12}     
     \ddot{a}_2&=\frac{GM'}{r^3}\,a_2(3\sin^2\al-1)+a_2(\la^2+\om^2)-2a_1\la\om +\left(\frac{5k_1}{n\ka_n}\frac{P_c}{\rho_c}\right)\frac{1}{a_2}-\frac{3GM}{\ka_n(1-\frac{n}{5})}\frac{a_2A_2}{2R^3} ,\\
\label{lageom13}
     \ddot{a}_3&=-\frac{GM'}{r^3}\,a_3+\left(\frac{5k_1}{n\ka_n}\frac{P_c}{\rho_c}\right)\frac{1}{a_3} -\frac{3GM}{\ka_n(1-\frac{n}{5})}\frac{a_3A_3}{2R^3} ,
\end{align}
\begin{align}
     \label{lageom14}
     \ddot{\psi}&=\dot{\la}=\frac{1}{a_1^2-a_2^2}\left[
     \frac{3GM'}{r^3}\,a_1a_2\,\sin 2\al
     -2(a_1\dot{a}_1-a_2\dot{a}_2)\la
     -2(\dot{a}_1a_2-a_1\dot{a}_2)\om\right] ,\\
\label{lageom15}     
     \ddot{\ga}&=\dot{\om}=\frac{1}{a_1^2-a_2^2}\left[\frac{3GM'}{2r^3}(a_1^2+a_2^2)\sin 2\al -2(a_1\dot{a}_1-a_2\dot{a}_2)\om -2(a_1\dot{a}_2-\dot{a}_1a_2)\la\right] ,
     \\
\label{lageom16}
     \ddot{r}&=r\dot{\phi}^2-\frac{G\M}{r^2}-\frac{3}{10}\frac{G\M}{r^4}
     \Bigl\{\ka_n\bigl[
     a_1^2(3\cos^2\al-1)+a_2^2(3\sin^2\al-1)-a_3^2\bigr] \nonumber \\
     &\hspace{0.5cm}
     +\ka_n'\bigl[a_1'^2(3\cos^2\al'-1)+a_2'^2(3\sin^2\al'-1)-a_3'^2\bigr]\Bigr\} ,\\
\label{lageom17}     
     \ddot{\phi}&=-2\,\frac{\dot{r}\dot{\phi}}{r}
     -\frac{3}{10}\frac{G\M}{r^5}\Bigl\{ (a_1^2-a_2^2)\ka_n\sin
     2\al+(a_1'^2-a_2'^2)\ka_n'\sin2\al'\Bigr\} ,
\end{align}
with $\M=M+M'$. 
The corresponding equations for $a_i',\ \psi'$ and $\ga'$ are obtained
by replacing unprimed variables by primed ones. Note that these
equations apply whenever $a_1\ne a_2$. If the two semi-major axes
$a_1$ and $a_2$ are equal, $a_1\equiv a_2$, some special care is
needed. In particular, Eqs. (\ref{lageom14}) and (\ref{lageom15}) are
ill defined, indicating that $\psi$ and $\ga$ are not independent
variables. In fact, setting $a_1\equiv a_2$, the Lagrangian
(\ref{lag1}) simplifies to  
\begin{align}
\label{lag2}
    L&=\frac{\mu}{2}\dot{r}^2+\frac{\mu
    r^2}{2}\dot{\phi}^2+\frac{GMM'}{r}+\frac{GMM'}{10r^3}\left[\ka_n(a_1^2-a_3^2)+\ka_n'(a_1'^2-a_3'^2)\right] +\frac{\ka_nM}{5}a_1^2\dot{\be}^2\nonumber \\
    &\hspace{0.5cm}
    +\frac{\ka_n'M'}{5}a_1'^2\dot{\be}'^2+\frac{\ka_nM}{10}(2\dot{a}_1^2+\dot{a}_3^2)+\frac{\ka_n'M'}{10}(2\dot{a}_1'^2+\dot{a}_3'^2)-k_1KM\rho_c^{1/n}
    \nonumber \\
    &\hspace{0.5cm}
    -k_1'K'M'\rho_c'^{1/n'} +\frac{3}{5-n}\frac{GM^2}{2R^3}\mc J+\frac{3}{5-n'}\frac{GM'^2}{2R'^3}\mc J',
\end{align}
where we introduced a new variable $\beta:=\ga-\psi$. The number of
degrees of freedom of the binary system is thus reduced from 12 to 8,
resulting in an enormeous simplification of the numerical calculations. Using
Eq. (\ref{Jdiv}), the equations of motion for binary systems with
axially symmetric stars become
\begin{align}
\label{lageom21}
    \ddot{a}_1&=\frac{GM'}{2r^3}\,a_1+a_1\dot{\be}^2+\left(\frac{5k_1}{\ka_nn}\frac{P_c}{\rho_c}\right)\frac{1}{a_1}-\frac{3GM}{\ka_n(1-\frac{n}{5})}\frac{a_1A_1}{2R^3} , \\
\label{lageom22}
    \ddot{a}_3&=-\frac{GM'}{r^3}\,a_3+\left(\frac{5k_1}{n\ka_n}\frac{P_c}{\rho_c}\right)\frac{1}{a_3}-\frac{3GM}{\ka_n(1-\frac{n}{5})}\frac{a_3A_3}{2R^3}     ,\\
\label{lageom23}
    \ddot{r}&=r\dot{\phi}^2-\frac{G\M}{r^2}-\frac{3}{10}\frac{G\M}{r^4}\Bigl\{\ka_n(a_1^2-a_3^2)+\ka_n'(a_1'^2-a_3'^2)\Bigr\} ,\\
\label{lageom24}    
    \ddot{\phi}&=-2\,\frac{\dot{r}\dot{\phi}}{r} ,\\
\label{lageom25}
    \ddot{\be}&=-2\frac{\dot{a}_1}{a_1}\,\dot{\be}.    
\end{align}
Later on we shall study the gravitational waves emitted by such
particular systems in detail, but first we shall derive the
Hamiltonian formalism for general Riemann-S binaries.
\begin{figure}[t]
\begin{center}
\psfrag{a}{$\al$}
\psfrag{g}{$\ga$}
\psfrag{p}{$\phi$}
\psfrag{y1}{$x_{\b1}$}
\psfrag{y2}{$x_{\b2}$}
\psfrag{x1}{$x_1$}
\psfrag{x2}{$x_2$}
\psfrag{M}{$M$}
\psfrag{M2}{$M'$}
\psfrag{com}{COM}
\includegraphics[width=7cm]{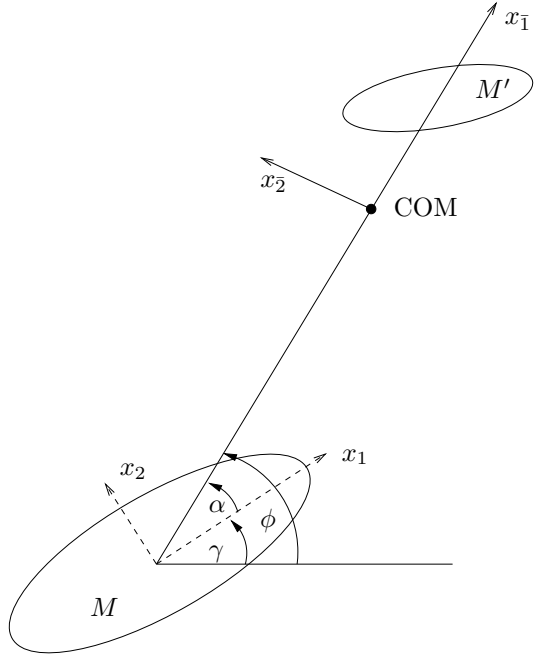}
\caption{The figure shows the coordinate systems used throughout
this paper. The $x_{\b 1}$-axes of the comoving system $\b \Sigma$
is chosen along the axes between the centers of $M$ and $M'$ and
includes with the $X-$axes of the inertial system the angle
$\phi$. Dotted lines represent the axes of the body-fixed system. COM
denotes the center of mass.} 
\end{center}
\end{figure}

\subsection{Riemann-S binary systems in Hamiltonian formalism}

In the previous section we have revisited the derivation of the
Lagrangian equations of motion for a Riemann-S binary. For the
following analysis it is, however, more suitable to apply the
Hamiltonian formulation of the problem. The generalized momenta
$p_i=\pa L/\pa\dot{q}_i$ derived from the Lagrangian (\ref{lag1}) 
are given by
\begin{align}
\label{hameom11}
    p_r&=\mu \dot{r},\qquad
    p_\phi=\mu r^2\dot{\phi}, \\
    p_{a_i}&=\frac{\ka_nM}{5}\dot{a}_i,\qquad
    p_\psi=\frac{\ka_nM}{5}\left[(a_1^2+a_2^2)\dot{\psi}-
    2a_1a_2\dot{\ga}\right],\qquad
    p_\ga=\frac{\ka_nM}{5}\left[
    (a_1^2+a_2^2)\dot{\ga}-2a_1a_2\dot{\psi}\right]. \nonumber
\end{align}
For $a_1\ne a_2$ it is possible to invert these equations in order to
express the generalized velocities in terms of generalized momenta:
\begin{align}
\label{hameom11b}   
    \dot{r}&=\frac{p_r}{\mu},\qquad
    \dot{\phi}=\frac{p_\phi}{\mu r^2},\qquad
    \dot{a}_i=\frac{5}{\ka_nM}\, p_{a_i},\\
    \dot{\ga}&=\frac{5}{\ka_nM}\left[\frac{a_1^2+a_2^2}{(a_1^2-a_2^2)^2}p_\ga
    +2\,\frac{a_1a_2}{(a_1^2-a_2^2)^2}\,p_\psi\right],\qquad
    \dot{\psi}=\frac{5}{\ka_nM}\left[
    \frac{a_1^2+a_2^2}{(a_1^2-a_2^2)^2}p_\psi
    +2\,\frac{a_1a_2}{(a_1^2-a_2^2)^2}p_\ga\right]. \nonumber  
\end{align}
The corresponding equations for $a_i',\ \psi'$ and $\ga'$   are
obtained by replacing unprimed variables by primed ones. For
$a_1\equiv a_2$ the generalized momenta $p_\psi$ and $p_\ga$ are not
independent variables, as can be easily seen from
Eqs. (\ref{hameom11}). This is reflected by the observation that the
inversion problem $p_i(\dot{q}_i)\to \dot{q}_i(p_i)$ is ill defined in
this case. We shall turn to this later on, but for now we focus on the
most general case, assuming $a_1\ne a_2$.  A Legendre transformation
of the Lagrangian (\ref{lag1}) leads to the Hamiltonian 
\begin{align}
\label{ham1}
     H&=\frac{p_r^2}{2\mu}+\frac{p_\phi^2}{2\mu
     r^2}+\frac{5}{2\ka_nM}\sum_{i=1}^3 p_{a_i}^2
     +\frac{5}{2\ka_n'M'}\sum_{i=1}^3 p_{a_i}'^2
     +\frac{5}{2\ka_nM}\frac{a_1^2+a_2^2}{(a_1^2-a_2^2)^2}(p_\ga^2+p_\psi^2)
     \nonumber \\
     &\hspace{0.5cm}
     +\frac{10}{\ka_nM}\frac{a_1a_2}{(a_1^2-a_2^2)^2}\,p_\ga p_\psi 
     +\frac{5}{2\ka_n'M'}\frac{a_1'^2+a_2'^2}{(a_1'^2-a_2'^2)^2}(p_\ga'^2+p_\psi'^2) +\frac{10}{\ka_n'M'}\frac{a_1'a_2'}{(a_1'^2-a_2'^2)^2}\,p_\psi'p_\ga'
     \nonumber \\
     &\hspace{0.5cm}
     +k_1KM\rho_c^{1/n}-\frac{3}{5-n}\frac{GM^2}{2R^3}\,\mc J
     +k_1'K'M'\rho_c'^{1/n'}-\frac{3}{5-n'}\frac{GM'^2}{2R'^3}\,\mc J'
     -\frac{G\M\mu}{r}
     \nonumber \\
     &\hspace{0.5cm}
     -\frac{G\M\mu\ka_n}{10r^3}
     \Bigl[a_1^2(3\cos^2\al-1)+a_2^2(3\sin^2\al-1)-a_3^2\Bigr]
     \nonumber \\
     &\hspace{0.5cm}
      -\frac{G\M\mu\ka_n'}{10r^3}
      \Bigl[a_1'^2(3\cos^2\al'-1)+a_2'^2(3\sin^2\al'-1)-a_3'^2\Bigr] . 
\end{align}
Once the Hamiltonian is given, the corresponding equations of motion
\begin{align*}
    \dot{q}_i=\frac{\pa H}{\pa p_i},\qquad
    \dot{p}_i=-\frac{\pa H}{\pa q_i}
\end{align*}
are easy to calculate using the relation for $\pa \mc J/\pa a_i$ derived
in the appendix. With the equations for the $\dot{q}_i$ already given
in Eqs. (\ref{hameom11b}) the remaining equations read
\begin{align}
\label{hameom12}
    \dot{p}_{a_1}&=\frac{5}{\ka_nM}\left[
    \frac{a_1(3a_2^2+a_1^2)}{(a_1^2-a_2^2)^3}(p_\ga^2+p_\psi^2)
    +2a_2\frac{a_2^2+3a_1^2}{(a_1^2-a_2^2)^3}\,p_\psi p_\ga\right]
    +\frac{G\M\mu}{5r^3}\ka_n a_1(3\cos^2\al-1) \nonumber \\
   &\hspace{0.5cm}
    +\frac{k_1M}{na_1}\frac{P_c}{\rho_c}
    -\frac{3GM^2}{5-n}\frac{a_1A_1}{2R^3} ,\\
\label{hameom13}
    \dot{p}_{a_2}&=-\frac{5}{\ka_nM}\left[
    \frac{a_2(3a_1^2+a_2^2)}{(a_1^2-a_2^2)^3}(p_\psi^2+p_\ga^2)
    +2a_1\frac{a_1^2+3a_2^2}{(a_1^2-a_2^2)^3}p_\psi p_\ga\right]
    +\frac{G\M\mu}{5r^3}\ka_na_2(3\sin^2\al-1) \nonumber \\
   &\hspace{0.5cm}
    +\frac{k_1M}{na_2}\frac{P_c}{\rho_c}
    -\frac{3GM^2}{5-n}\frac{a_2A_2}{2R^3} , \\
\label{hameom14}
    \dot{p}_{a_3}&=\frac{k_1M}{na_3}\frac{P_c}{\rho_c}
    -\frac{3GM^2}{5-n}\frac{a_3A_3}{2R^3}-\frac{G\M\mu}{5r^3}\,
    \ka_na_3   ,\\
\label{hameom15}
    \dot{p}_\psi&=0 ,\\
\label{hameom16}
    \dot{p}_\ga&=\frac{3}{10}\frac{G\M\mu}{r^3}\,\ka_n(a_1^2-a_2^2)\sin 2\al
    ,\\
\label{hameom17}
    \dot{p}_r&=-\frac{G\M\mu}{r^2}+\frac{p^2_\phi}{\mu r^3}-\frac{3}{10}\frac{G\M\mu\ka_n}{r^4}
    \Bigl[ a_1^2(3\cos^2\al-1)+a_2^2(3\sin^2\al-1) -a_3^2\Bigr]
    \nonumber \\
    &\hspace{0.5cm}
    -\frac{3}{10}\frac{G\M\mu\ka_n'}{r^4}\Bigl[a_1'^2(3\cos^2\al'-1)
    +a_2'^2(3\sin^2\al'-1) -a_3'^2\Bigr] ,\\
\label{hameom18}
     \dot{p}_\phi&=-\frac{3}{10}\frac{G\M\mu}{r^3}\left[
     \ka_n(a_1^2-a_2^2)\sin 2\al +\ka_n'(a_1'^2-a_2'^2)\sin 2\al'\right].       \end{align} 
Together with the corresponding equations for the primed variables
this system of differential equations describes the dynamics of the
binary completely.\\

Let us now assume $a_1\equiv a_2,\ a_1'\equiv a_2'$. As we have already
pointed out in this case the dynamical equations (\ref{hameom11b}) and
(\ref{hameom12})-(\ref{hameom18}) do not apply. The generalized momenta
derived from Lagrangian (\ref{lag2}) read
\begin{align}
\label{hameom21}
       p_{a_1}&=\frac{2}{5}\ka_nM\dot{a}_1,\qquad
       p_{a_3}=\frac{\ka_nM}{5}\dot{a_3},\qquad
       p_{\be}=\frac{2}{5}\ka_nMa_1^2\dot{\be}, \nonumber\\
       p_r&=\mu \dot{r},\qquad
       p_\phi=\mu r^2\dot{\phi} ,
\end{align}
which can be easily inverted, giving
\begin{align}
\label{hameom21b}
       \dot{a}_1&=\frac{5}{2\ka_nM}\,p_{a_1},\qquad
       \dot{a}_3=\frac{5}{\ka_nM}\,p_{a_3},\qquad
       \dot{\be}=\frac{5}{2\ka_nM}\frac{p_\be}{a_1^2} , \\
       \dot{r}&=\frac{p_r}{\mu},\qquad
       \dot{\phi}=\frac{p_\phi}{\mu r^2} . \nonumber
\end{align}
As before a Legendre transformation of the Lagrangian (\ref{lag2})
leads to the corresponding Hamiltonian
\begin{align}
\label{ham2}
    H&=\frac{5}{2\ka_nM}\left[\frac{p_{a_1}^2}{2}+p_{a_3}^2
    +\frac{p_\be^2}{2a_1^2}\right]
    +\frac{5}{2\ka_n'M'}\left[\frac{p_{a_1}'^2}{2}+p_{a_3}'^2
    +\frac{p_\be'^2}{2a_1'^2}\right] +\frac{p_r^2}{2\mu}
    +\frac{p_\phi^2}{2\mu r^2}
     +k_1KM\rho_c^{1/n} \nonumber \\
    &\hspace{0.5cm}
     -\frac{3GM^2}{5-n}\frac{\mc J}{2R^3}
    +k_1'K'M'\rho_c'^{1/n'} -\frac{3GM'^2}{5-n'}\frac{\mc J'}{2R'^3}
    -\frac{G\M\mu}{r}-\frac{G\M\mu}{10
    r^3}\left[\ka_n(a_1^2-a_3^2)+\ka_n'(a_1'^2-a_3'^2)\right]. 
\end{align}
Note that $\be,\be'$ and $\phi$ are cyclic variables, i.e. the
corresponding generalized momenta are constants. The Hamiltonian
equations are then given by Eqs. (\ref{hameom21b}) and
\begin{align}
\label{hameom22}
    \dot{p}_{a_1}&=\frac{5}{2\ka_nM}\frac{p_\be^2}{a_1^3}+\frac{G\M\mu\ka_n}{5r^3}\,a_1+2\,\frac{k_1M}{na_1}\frac{P_c}{\rho_c}-\frac{3GM^2}{5-n}\frac{a_1A_1}{R^3} , \\
\label{hameom23}
    \dot{p}_{a_3}&=-\frac{G\M\mu\ka_n}{5r^3}\,a_3
    +\frac{k_1M}{na_3}\frac{P_c}{\rho_c}
    -\frac{3GM^2}{5-n}\frac{a_3A_3}{2R^3} , \\
\label{hameom24}
   \dot{p}_r&=-\frac{G\M\mu}{r^2}+\frac{p_\phi^2}{\mu
   r^3}-\frac{3}{10}\frac{G\M\mu}{r^4}
   \left[\ka_n(a_1^2-a_3^2)+\ka_n'(a_1'^2-a_3'^2)\right] , \\
\label{hameom25}
   \dot{p}_\phi&=\dot{p}_\be=0 ,        
\end{align}
and, of course, the corresponding equations for $p_{a_i}'$ and $\be'$.

\section{Leading order radiation reaction in Riemann-S binaries}

On the Newtonian level, tidally coupled binaries form a conservative
system. However,
according to the theory of General Relativity binary systems loose
energy due to the emission of gravitational waves, the first
nonvanishing dissipative terms appearing at 2.5 post-Newtonian approximation.
The following section is devoted to the calculation of the leading
order radiation reaction terms for Riemann-S binaries. This can be
done by virtue of the Burke-Thorne radiation reaction potential
$\Phi_{reac}$, given e.g. in \cite{MTW},
\begin{align}
\label{reacpot}
     \Phi_{reac}=\frac{G}{5c^5} \Ib_{\b a\b b}^{(5)} x_{\b a}x_{\b b}.
\end{align}
As before the $x_{\b a}$ denote coordinates in the corotating
coordinate frame.  Thus the computation of the radiation reaction
potential reduces in principle to the calculation of the 5th time
derivative of the STF mass quadrupole tensor \textit{in the corotating
system}. The mass quadrupole tensor being additive at Newtonian order,
it is possible to consider orbital and stellar contributions
separately.

\subsection{Time derivatives of the stellar mass quadrupole tensor}

According to Eq. (\ref{reacpot}) the leading order gravitational wave
emission is governed by the time variations of the STF mass quadrupole
tensor. Thus even an isolated, oscillating star represents a source of
gravitational waves. In a binary system tidal coupling between stellar
and orbital degrees of freedom leads to a change in the gravitational
wave pattern. Though the orbital contribution in general clearly dominates the
gravitational waveforms emitted by the binary, the stellar
contributions cannot be neglected.
\\
Let us consider a coordinate transformation such that the origin of
the corotating system is centered at $M$. The relation between
the coordinates $X_a$ of an inertial system centered at $M$, the
coordinates $x_a$ of the body-framed system and the coordinates $x_{\b a}$
can be read off from Fig. 1,
\begin{align}
\label{trafo}
    x_a=T_{ab}(\ga)X_b,\qquad
    x_{\b a}=T_{\b ab}(\phi)X_b,
\end{align}
where
\begin{align}
   T_{ab}(\phi)=\left(\begin{array} {ccc}
   \cos\phi &\sin\phi & 0 \\
   -\sin\phi & \cos\phi & 0 \\
   0 & 0 & 1 \end{array}\right).
\end{align}
As has already been  mentioned, the STF mass quadrupole tensor
takes a particular simple form in the body-fixed system,
\begin{align}
     \Ib_{ab}=I_{ab}-\frac{1}{3}\delta_{ab}I_{cc},
     \qquad
     I_{ab}=\int \rho x_a x_b dV=\frac{\ka_nM}{5}a_a^2\delta_{ab}.
\end{align}
To calculate the time derivatives of the mass quadrupole tensor in the
corotating system we could in a first step 
consider the time derivatives of the star's mass quadrupole tensor in
the inertial system, the later one being related to $\Ib_{ab}$ by the
transformation
\begin{align*}
     \Ib^{(IS)}_{\al\be}=T^\dag(\ga)_{\al i}T^\dag_{\be j}(\ga)\Ib_{ij},
\end{align*}
and only in a second step apply the transformation into the corotating
system according to
\begin{align}
     \Ib^{(5)}_{\b a\b b}=T_{\b a\al}(\phi)T_{\b b\be}(\phi)\Ib^{(5,IS)}_{\al\be}.
\end{align}
This calculation is straightforward, but
successively inserting the Newtonian equations of motion leads to
rather complicated expressions. In our approach another strategy is
more suitable. 
Following Lai and Shapiro (see also \cite{Chandrasekhar:1970}) we combine
the steps mentioned above according to
\begin{align}
\label{reac1}
    \Ib^{(5)}_{\ba\bb}&=
        T_{\ba\al}(\phi)T_{\bb\be}(\phi)\frac{d^5}{dt^5}
        \left[T^\dag_{\al i}(\ga)T^\dag_{\be j}(\ga)
        \Ib_{ij}\right] \nonumber \\
    &=\sum_{m=0}^5\left(\begin{array} {c} 5 \\ m \end{array}\right)
        \left[\frac{d^{5-m}}{dt^{5-m}}\Ib_{ij}\right]
        \sum_{p=0}^m \left(\begin{array} {c} m \\ p \end{array}\right)
        \left[T_{\ba\al}(\phi)\frac{d^{m-p}}{dt^{m-p}}T^\dag_{\al
        i}(\ga)\right]
        \left[T_{\bb\be}(\phi)\frac{d^p}{dt^p}T^\dag_{\be j}(\ga)\right]
        \nonumber \\
    &=\sum_{m=0}^5 \sum_{p=0}^m \left(\begin{array} {c}5 \\ m
        \end{array}\right)
        \left(\begin{array} {c} m \\ p \end{array}\right)
        \left[\frac{d^{5-m}}{dt^{5-m}}\Ib_{ij}\right]
        R^{m-p}_{\ba i}R^p_{\bb j},
\end{align}
where
\begin{align*}
    R^p_{\ba i}:=T_{\ba \al}(\phi)\frac{d^p}{dt^p}T^\dag_{\al i}(\ga).
\end{align*}
The calculation of the matrices $R^p_{\ba i}$ is straightforward and 
given in the Appendix. However, the resulting expressions for $\Ib^{(5)}_{\ba\bb}$ being rather complicated, we should imply further assumptions
on the internal stellar motion. In particular, it is reasonable to consider all
internal velocities and accelerations to be small, thus applying the
quasi-static approximation to the stellar degrees of freedom. This
strategy was already followed by Lai and Shapiro \cite{Lai:1995a}, but they
applied the quasi-static assumption to the orbital motion, too. While
this is justified for circular orbits or if one is interested in a few
cycles only, for elliptical orbits this leads  to a growing phase error
in the gravitational waveforms.  Since we are interested in the
long-term evolution of the system we shall apply the quasi-static
approximation to the stars only. 
Neglecting all terms of
order $O(\ddot a_i)$ and $O(\ddot{\Omega})$, keeping only terms linear
in $\dot{a}_i$ and $\dot{\Omega}$ and using that $|\dot{a}_i|\ll |\om
a_i|$ , Eq. (\ref{reac1}) simplifies to
\begin{align}
\label{reac2}
   \Ib^{(5)}_{\ba\bb}&=\Ib_{ij}
   \sum_{p=0} ^5 \left(\begin{array} {c} 5 \\ p\end{array}\right)
   R_{\ba i}^pR_{\bb j}^{5-p}
   +\dot{\Ib}_{ij}\sum_{p=0}^4
   \left(\begin{array} {c} 4 \\ p \end{array}\right)
   R_{\ba i}^p R_{\bb j}^{4-p}.
\end{align}
Note that $\ddot{\Ib}_{ab}\approx 0$ in this approximation. After some
algebra we end up with
\begin{align}
\label{reac3}
    \Ib^{(5)}_{\ba\bb}&=(I_{11}-I_{22})\left[
    16\om^5\,\left(\begin{array} {ccc}
    \sin 2\al & \cos 2\al & 0 \\
    \cos 2\al & -\sin 2\al & 0\\
    0 & 0 & 0 \end{array}\right) -
    80\om^3\dot{\om}\,\left(\begin{array} {ccc}
   - \cos 2\al & \sin 2\al & 0 \\
    \sin 2\al & \cos 2\al & 0 \\
    0 & 0 & 0 \end{array}\right)\right] \nonumber \\
    &+40\om^4(\dot{I}_{11}-\dot{I}_{22})\left(\begin{array} {ccc}
    \cos 2\al & -\sin 2\al & 0 \\
    -\sin 2\al & -\cos 2\al & 0 \\
    0 & 0 & 0 \end{array}\right)
\end{align}
It is remarkable that  the components of $\Ib^{(5)}_{\ba\bb}$ are 
nonvanishing only for $a_1\ne a_2$. In other words, the quasi-static
approximation does not allow for gravitational wave emission of an
isolated, axially symmetric, polytropic star\footnote{Without the
quasi-approximation there exist nonvanishing contributions to
$\Ib^{(5)}_{\ba\bb}$  for axially symmetric polytropes, too.}.

\subsection{The orbital contribution to $\Ib^{(5)}_{\ba\bb}$}

As has been outlined before, the quasi-static approximation does not
apply to the orbital motion in general. To calculate
$\Ib^{(5),orb}_{\ba\bb}$ we compute $\Ib^{(5),orb}_{ij}$ in the
inertial system by successively inserting the Newtonian equations of
motion
\begin{align}
\label{newt}
    \ddot{r}=-\frac{G\M}{r^2}+r\dot{\phi}^2,\qquad
    \ddot{\phi}=-2\frac{\dot{r}\dot{\phi}}{r}.
\end{align}
More precisely, one should insert the equations of motion of the
tidally coupled system (cf. Eqs. (\ref{lageom16}) and
(\ref{lageom17})). This would complicate the problem
enormeously. However, these 
corrections, being of higher order in $1/r$, can easily be
neglected.\\
In the inertial system the nonvanishing components of
$\Ib^{(orb)}_{ij}$ read
\begin{align}
\label{orbit}
     \Ib^{(orb)}_{11}=\frac{\mu r^2}{6}(1+3\,\cos 2\phi),\qquad
     \Ib^{(orb)}_{22}=\frac{\mu r^2}{6}(1-3\,\cos 2\phi),\qquad
     \Ib^{(orb)}_{33}=-\frac{\mu r^2}{3},\qquad
     \Ib^{(orb)}_{12}=\frac{\mu r^2}{2}\,\sin 2\phi.
\end{align}
Using Eqs. (\ref{newt}) to calculate $\Ib^{(5)}_{ij}$ in the inertial
system and finally transfering to the corotating system centered at
the center of mass according to
\begin{align}
     \Ib^{(5,orb)}_{\ba\bb}&=T_{\ba\al}(\phi)T_{\bb\be}(\phi)\Ib^{(5),(orb,IS)}_{\al\be},
\end{align}
we end up with
\begin{align}
\label{reac4}
     \Ib^{(5),orb}_{\b1\b1}&=-\frac{8G\M\mu}{3}\frac{\dot{r}}{r^4}\left[4\,\frac{G\M}{r}+3\dot{r}^2+18r^2\dot{\phi}^2\right],\nonumber \\
     \Ib^{(5),orb}_{\b2\b2}&=\frac{2G\M\mu}{3}\frac{\dot{r}}{r^4}\left[8\,\frac{G\M}{r}+6\dot{r}^2+81r^2\dot{\phi}^2\right], \nonumber \\
     \Ib^{(5),orb}_{\b3\b3}&=\frac{2G\M\mu}{3}\frac{\dot{r}}{r^4}\left[8\,\frac{G\M}{r}+6\dot{r}^2-9r^2\dot{\phi}^2\right], \nonumber \\
     \Ib^{(5),orb}_{\b1\b2}&=-4\,\frac{G\M\mu}{r^3}\,\dot{\phi}
     \left[\frac{2G\M}{r}+9\dot{r}^2-6r^2\dot{\phi}^2\right].
\end{align}

\subsection{Leading order gravitational radiation -- general case}

In a recent paper we discussed the dynamics and gravitational wave
emission of a binary system consisting of a rotating, oscillating
dusty disk and a point mass \cite{Hansen:2005}. There we incorporated the leading
order radiation reaction into the Hamiltonian equations by adding a
dissipative part to the Hamiltonian. The Hamiltonian equations can
then be applied in the usual way. Here we choose an alternative
approach, leaving the Lagrangian or Hamiltonian unchanged but modifying
the Euler-Lagrangian equations according to
\begin{align}
\label{lagdiss}
     \frac{d}{dt}\frac{\pa L}{\pa \dot{q}_i}=\frac{\pa L}{\pa q_i}+\F_{q_i}.
\end{align}
The generalized dissipative forces $\F_{q_i}$  are calculated from the
energy dissipation rate
\begin{align}
\label{diss}
     \W=-\int \vec v\cdot\nabla \Phi_{reac}\rho \,dV
\end{align}
as $\F_{q_i}=\frac{\pa W}{\pa \dot{q}_i}$. Let us consider the
contribution $\W_M$ of star $M$ to the energy dissipation
rate of the binary. The velocity $\vec v$ of a fluid element of $M$
can be separated according to $\vec v=\vec u+\vec u_{orb}$, where
\begin{align*}
    \vec u=\left(\frac{a_1}{a_2}\la-\om\right)x_2\vec
    e_1+\left(-\frac{a_2}{a_1}\la+\om\right)x_1\vec e_2 +
    \frac{\dot{a}_1}{a_1}x_1\vec e_1+\frac{\dot{a}_2}{a_2}x_2\vec e_2
    +\frac{\dot{a}_3}{a_3}x_3\vec e_3
\end{align*}
is the velocity of a fluid element relative to the center of $M$ and
\begin{align}
     \vec u_{orb}=-\dot{r}_M\vec e_{\b1}-r_M\dot{\phi}\vec e_{\b2}
\end{align}
is the orbital velocity of the star's center of mass. As before
the $\vec e_i$ denote the unit vectors in the body-fixed coordinate
system, while the $\vec e_{\ba}$ represent unit vectors in
the corotating system.
As one can easily read off from Fig. 1, the coordinates are related
according to
\begin{align}
\label{trafo2}
   x_{\b1}=x_1\cos\al +x_2\sin\al-r_M ,\qquad
   x_{\b2}=-x_1\sin\al+x_2\cos\al,\qquad
   x_{\b3}=x_3.
\end{align}
Inserting this into Eq. (\ref{diss}) and using $\int\rho
x_ix_jdV=\frac{\ka_nM}{5}a_i^2\delta_{ij}$ the contribution of the
star $M$ to the energy dissipation rate yields
\begin{align}
\label{diss2}
    \W_M&=-\frac{2G}{5c^5}\frac{\ka_nM}{5}\left[a_1\dot{a}_1\bigl(
    \Ib^{(5)}_{\b1\b1}\cos^2\al+\Ib^{(5)}_{\b2\b2}\sin^2\al-
    \Ib^{(5)}_{\b1\b2}\sin 2\al\bigr)
    +a_2\dot{a}_2\bigl(\Ib^{(5)}_{\b1\b1}\sin^2\al+\Ib^{(5)}_{\b2\b2}\cos^2\al
    +\Ib^{(5)}_{\b1\b2}\sin 2\al\bigr) \right. \nonumber \\
    &\hspace{2cm}
    +\left. \om(a_1^2-a_2^2)\bigl(\Ib^{(5)}_{\b1\b2}\cos 2\al
    +\frac{1}{2}(\Ib^{(5)}_{\b1\b1}-\Ib^{(5)}_{\b2\b2})\sin 2\al\bigr)
    +\Ib^{(5)}_{\b3\b3}a_3\dot{a}_3\right] \nonumber \\
    &\hspace{0.5cm}
    -\frac{2GM}{5c^5}\left[r_M\dot{r}_M\Ib^{(5)}_{\b1\b1}+r_M^2\dot{\phi}\,\Ib^{(5)}_{\b1\b2}\right] .
\end{align}
The corresponding contribution of $M'$ is obtained by replacing
unprimed quantities by primed ones. Note that according to Fig. 1
the orbital velocity of $M'$ is given by
$\vec u_{orb}'=\dot{r}_M'\vec e_{\b1} +r_M'\dot{\phi}\,\vec
e_{\b2}$. Adding up both contributions and using
$Mr_M\dot{r}_M+M'r_M'\dot{r}_M'=\mu r\,\dot{r}$,  the energy dissipation
rate of the binary system reads
\begin{align}
\label{disstotal}
     \W&=-\frac{2G}{5c^5}\frac{\ka_nM}{5}\left[a_1\dot{a}_1\bigl(
    \Ib^{(5)}_{\b1\b1}\cos^2\al+\Ib^{(5)}_{\b2\b2}\sin^2\al-
    \Ib^{(5)}_{\b1\b2}\sin 2\al\bigr)
    +a_2\dot{a}_2\bigl(\Ib^{(5)}_{\b1\b1}\sin^2\al+\Ib^{(5)}_{\b2\b2}\cos^2\al
    +\Ib^{(5)}_{\b1\b2}\sin 2\al\bigr) \right. \nonumber \\
    &\hspace{2cm}
    +\left. \om(a_1^2-a_2^2)\bigl(\Ib^{(5)}_{\b1\b2}\cos 2\al
    +\frac{1}{2}(\Ib^{(5)}_{\b1\b1}-\Ib^{(5)}_{\b2\b2})\sin 2\al\bigr)
    +\Ib^{(5)}_{\b3\b3}a_3\dot{a}_3\right] \nonumber \\
    &-\frac{2G}{5c^5}\frac{\ka_n'M'}{5}\left[a_1'\dot{a}_1'\bigl(
    \Ib^{(5)}_{\b1\b1}\cos^2\al'+\Ib^{(5)}_{\b2\b2}\sin^2\al'-
    \Ib^{(5)}_{\b1\b2}\sin 2\al'\bigr)
    +a_2'\dot{a}_2'\bigl(\Ib^{(5)}_{\b1\b1}\sin^2\al'+
    \Ib^{(5)}_{\b2\b2}\cos^2\al'
    +\Ib^{(5)}_{\b1\b2}\sin 2\al'\bigr) \right. \nonumber \\
    &\hspace{2cm}
    +\left. \om(a_1'^2-a_2'^2)\bigl(\Ib^{(5)}_{\b1\b2}\cos 2\al'
    +\frac{1}{2}(\Ib^{(5)}_{\b1\b1}-\Ib^{(5)}_{\b2\b2})\sin 2\al'\bigr)
    +\Ib^{(5)}_{\b3\b3}a_3'\dot{a}_3'\right] \nonumber \\
    &-\frac{2G\mu}{5c^5}(r\,\dot{r}\Ib^{(5)}_{\b1\b1}+r^2\dot{\phi}\,\Ib^{(5)}_{\b1\b2})  \quad .
\end{align}
It is now straightforward to derive the generalized forces
\begin{align}
\label{forces1}
     \F_{a_1}&=-\frac{2G}{5c^5}\frac{\ka_nM}{5}\bigl[
     \Ib^{(5)}_{\b1\b1}\cos^2\al+\Ib^{(5)}_{\b2\b2}\sin^2\al-
     \Ib^{(5)}_{\b1\b2}\sin 2\al\bigr]a_1 ,\nonumber \\
     \F_{a_2}&=-\frac{2G}{5c^5}\frac{\ka_nM}{5}\bigl[
     \Ib^{(5)}_{\b1\b1}\sin^2\al+\Ib^{(5)}_{\b2\b2}\cos^2\al+
     \Ib^{(5)}_{\b1\b2}\sin 2\al\bigr]a_2 ,\nonumber \\
     \F_{a_3}&=-\frac{2G}{5c^5}\frac{\ka_nM}{5}\Ib^{(5)}_{\b3\b3}a_3
     ,\nonumber \\
     \F_\ga&=-\frac{2G}{5c^5}\frac{\ka_nM}{5}(a_1^2-a_2^2)
     \bigl(\Ib^{(5)}_{\b1\b2}\cos
     2\al+\frac{1}{2}(\Ib^{(5)}_{\b1\b1}-\Ib^{(5)}_{\b2\b2})\sin
     2\al\bigr) ,\nonumber \\
     \F_\psi&=0 ,\nonumber \\
     \F_r&=-\frac{2G\mu}{5c^5}\Ib^{(5)}_{\b1\b1}\,r ,\nonumber \\
     \F_\phi&=-\frac{2G\mu}{5c^5}\,\Ib^{(5)}_{\b1\b2} r^2,
\end{align}
the corresponding generalized forces for $M'$ being obtained by
replacing unprimed variables by primed ones. The Euler-Lagrangian
equations governing the dynamics of the system read now 
\begin{align}
\label{disslag}
    \ddot{a}_1&= [ \text{Eq.\  (\ref{lageom11})} ]
    -\frac{2G}{5c^5}(\Ib^{(5)}_{\b1\b1}\cos^2\al
    +\Ib^{(5)}_{\b2\b2}\sin^2\al -\Ib^{(5)}_{\b1\b2}\sin 2\al)a_1 ,\nonumber\\
    \ddot{a}_2&=[ \text{Eq.\  (\ref{lageom12})} ]
    -\frac{2G}{5c^5}(\Ib^{(5)}_{\b1\b1}\sin^2\al+\Ib^{(5)}_{\b2\b2}\cos^2\al+\Ib^{(5)}_{\b1\b2}\sin 2\al)a_2 ,\nonumber \\
    \ddot{a}_3&= [ \text{Eq.\  (\ref{lageom13})} ]
    -\frac{2G}{5c^5}\Ib^{(5)}_{\b3\b3}a_3 ,\nonumber \\
    \dot{\la}&=\left(\frac{a_1}{a_2}-\frac{a_2}{a_1}\right)^{-1}
    \left[
    -2\left(\frac{\dot{a}_1}{a_2}-\frac{\dot{a}_2}{a_1}\right)\la
    -2\left(\frac{\dot{a}_1}{a_1}-\frac{\dot{a}_2}{a_2}\right)\om
    +\frac{3GM'}{r^3}\sin 2\al  \right. \nonumber \\
    &\hspace{2.5cm}
    \left.-\frac{2G}{5c^5}\bigl(
    2\Ib^{(5)}_{\b1\b2}\cos 2\al
    +(\Ib^{(5)}_{\b1\b1}-\Ib^{(5)}_{\b2\b2})\sin 2\al\bigr)\right]
    ,\nonumber \\
     \dot{\om}&=\left(\frac{a_1}{a_2}-\frac{a_2}{a_1}\right)^{-1}
     \left[
     -2\left(\frac{\dot{a}_1}{a_1}-\frac{\dot{a}_2}{a_2}\right)\la
     -2\left(\frac{\dot{a}_1}{a_2}-\frac{\dot{a}_2}{a_1}\right)\om
     \right. \nonumber \\
     &\hspace{2.5cm}+ \left.
     \left(\frac{a_1}{a_2}+\frac{a_2}{a_1}\right)
     \left\{\frac{3}{2}\frac{GM'}{r^3}\,\sin 2\al
     -\frac{2G}{5c^5}(\Ib^{(5)}_{\b1\b2}\cos 2\al
     +\frac{1}{2}(\Ib^{(5)}_{\b1\b1}-\Ib^{(5)}_{\b2\b2})\sin 2\al)
     \right\}
     \right] ,\nonumber \\
     \ddot{r}&=
     r\dot{\phi}^2-\frac{G\M}{r^2}-\frac{3}{10}\frac{G\M}{r^4}
     \left[
     \ka_n\bigl(a_1^2(3\cos^2\al-1)+a_2^2(3\sin^2\al-1)-a_3^2\bigr)
     \right. \nonumber \\
     &\hspace{0.5cm} \left.
     +\ka_n'\bigl(a_1'^2(3\cos^2\al'-1)+a_2'^2(3\sin^2\al'-1)-a_3'^2\bigr)
     \right]
     -\frac{2G}{5c^5}\,\Ib^{(5)}_{\b1\b1}\, r   \quad  ,\nonumber \\
     \ddot{\phi}&=-2\,\frac{\dot{r}\dot{\phi}}{r}-\frac{3G\M}{10r^5}
     \left[
     \ka_n(a_1^2-a_2^2)\sin 2\al +\ka_n'(a_1'^2-a_2'^2)\sin 2\al'
     \right]
     -\frac{2G}{5c^5}\,\Ib^{(5)}_{\b1\b2}\quad .
\end{align}
Together with the corresponding equations for $M'$ these equations describe
the evolution of the binary system including leading order
gravitational reaction. Note that  in  Eqs. (\ref{disslag}) all
contributions to the leading order gravitational reaction are included. 
The quasi-static approximation for the stellar degrees of freedom
enters into the explicit calculation of the time
derivatives of the STF mass quadrupole tensor.

\subsection{Specializing to binary systems with $a_1=a_2, a_1'=a_2'$}

In the previous section we derived the Lagrangian equations of motion
for arbitrary Riemann-S binaries. From now on we shall impose an
additional constraint on the stellar degrees of freedom, requiring
$a_1=a_2,\ a_1'=a_2'$. On Newtonian level the equations of motion
governing the dynamics of those systems have been derived in section
2. Now let us consider the radiation reaction part of the equations of
motion. In principle we follow the same strategy as in the previous
section. The velocity of a fluid element of star $M$ relative to the
star's center reads now
\begin{align*}
   \vec u=-\dot{\be}x_2\vec e_1+\dot{\be}x_1\vec
   e_2+\frac{\dot{a}_1}{a_1}(x_1\vec e_1+x_2\vec e_2)
   +\frac{\dot{a}_3}{a_3}x_3\vec e_3,
\end{align*}
and the orbital motion is $\vec
u_{orb}=-\dot{r}_M\vec e_{\b1}-r_M\dot{\phi}\vec e_{\b2}$. Calculating
$\W_M$ and $\W_M'$ according to Eq. (\ref{diss}), the gravitational energy
dissipation rate of the binary system yields
\begin{align}
\label{dissspec}
     \W&=-\frac{2G}{5c^5}\frac{\ka_nM}{5}
     \left[
     a_1\dot{a}_1(\Ib^{(5)}_{\b1\b1}+\Ib^{(5)}_{\b2\b2})+
     a_3\dot{a}_3\Ib^{(5)}_{\b3\b3}
     \right]
     -\frac{2G}{5c^5}\frac{\ka_n'M'}{5}
     \left[a_1'\dot{a}_1'(\Ib^{(5)}_{\b1\b1}+\Ib^{(5)}_{\b2\b2})
     +a_3'\dot{a}_3'\Ib^{(5)}_{\b3\b3}\right] \nonumber \\
     &\hspace{0.5cm}
     -\frac{2G\mu}{5c^5}(r\dot{r}\,\Ib^{(5)}_{\b1\b1} +r^2\dot{\phi}\,\Ib^{(5)}_{\b1\b2}).
\end{align}
The generalized dissipative forces take on a particular simple form:
\begin{align}
\label{forces2}
      \F_{a_1}&=-\frac{2G}{5c^5}\frac{\ka_nM}{5}(\Ib^{(5)}_{\b1\b1}+\Ib^{(5)}_{\b2\b2})a_1=\frac{2G}{5c^5}\frac{\ka_nM}{5}\Ib^{(5)}_{\b3\b3}a_1 , \nonumber \\
      \F_{a_3}&=-\frac{2G}{5c^5}\frac{\ka_nM}{5}\Ib^{(5)}_{\b3\b3}a_3
      ,\nonumber \\
      \F_\be&=0 ,\nonumber \\
      \F_r&=-\frac{2G\mu}{5c^5}\,\Ib^{(5)}_{\b1\b1}r ,\nonumber \\
      \F_\phi&=-\frac{2G\mu}{5c^5}\,\Ib^{(5)}_{\b1\b2}r^2.
\end{align}
So the Lagrangian equations of motion including leading order
radiation reaction terms are given by
\begin{align}
\label{disslagspec}
    \ddot{a}_1&=\frac{GM'}{2r^3}\,a_1+a_1\dot{\be}^2+
    \left(\frac{5k_1}{\ka_nn}\frac{P_c}{\rho_c}\right)\frac{1}{a_1}
    -\frac{3GM}{\ka_n(1-\frac{n}{5})}\frac{a_1A_1}{2R^3}
    +\frac{G}{5c^5}\Ib^{(5)}_{\b3\b3}a_1
    ,\nonumber \\
    \ddot{a}_3&=-\frac{GM'}{r^3}a_3
    +\left(\frac{5k_1}{\ka_nn}\frac{P_c}{\rho_c}\right)\frac{1}{a_3}
    -\frac{3GM}{\ka_n(1-\frac{n}{5})}\frac{a_3A_3}{2R^3}
    -\frac{2G}{5c^5}\,\Ib^{(5)}_{\b3\b3}a_3
    ,\nonumber \\
    \ddot{\be}&=-2\,\frac{\dot{a}_1}{a_1}\,\dot{\be}
    ,\nonumber \\
    \ddot{r}&=r\dot{\phi}^2-\frac{G\M}{r^2}
    -\frac{3G\M}{10r^4}\left[\ka_n(a_1^2-a_3^2)+\ka_n'(a_1'^2-a_3'^2)\right]
    -\frac{2G}{5c^5}\,\Ib^{(5)}_{\b1\b1}\,r
    ,\nonumber \\
    \ddot{\phi}&=-2\,\frac{\dot{r}\dot{\phi}}{r}-\frac{2G}{5c^5}\,\Ib^{(5)}_{\b1\b2}.
\end{align}
At this point we should emphasize that in the quasi-static
approximation the stellar contribution to $\Ib^{(5)}_{\ba\bb}$
vanishes, as can be easily seen from Eq. (\ref{reac3}). This means,
in that approximation only the time varying orbital mass quadrupole
tensor gives rise to the emission of gravitational waves. However,
coupling the internal dynamics of the stars with the orbital dynamics
both, the orbital dynamics as well as the gravitational waveforms are
affected by the internal dynamics.
\\
For our purposes it is more suitable to describe the dynamics of the
binary system in Hamiltonian formalism. In the presence of dissipative
forces the well-known Hamiltonian equations are modified according to
\begin{align}
\label{hamdiss}
     \dot{q}_i=\frac{\pa H}{\pa p_i},\qquad
     \dot{p}_i=-\frac{\pa H}{\pa q_i} +\F_{q_i},
\end{align}
where the Hamiltonian $H$ is given by Eq. (\ref{ham1}) and Eq.
(\ref{ham2}), respectively. The generalized momenta are defined in the
usual way according to $p_i=\pa
L/\pa\dot{q}_i$.  The Hamiltonian equations derived in section
2.2. (cf. Eqs. 
(\ref{hameom22})-(\ref{hameom25})) are thus altered according to 
\begin{align}
\label{hamdiss2}
     \dot{p}_{a_1}&=\frac{5}{2\ka_nM}\frac{p_\be^2}{a_1^3}
     +\frac{G\M\mu\ka_n}{5r^3}\,a_1
     +2\frac{k_1M}{na_1}\frac{P_c}{\rho_c}
     -\frac{3GM^2}{5-n}\frac{a_1A_1}{R^3}
     +\frac{2G}{5c^5}\frac{\ka_nM}{5} \Ib^{(5)}_{\b3\b3}a_1
     \ , \nonumber \\
     \dot{p}_{a_3}&=-\frac{G\M\mu\ka_n}{5r^3}\,a_3
     +\frac{k_1M}{na_3}\frac{P_c}{\rho_c}
     -\frac{3GM^2}{5-n}\frac{a_3A_3}{2R^3}
     -\frac{2G}{5c^5}\frac{\ka_nM}{5}\,\Ib^{(5)}_{\b3\b3}a_3
     \ , \nonumber \\
     \dot{p}_\be&=0, \nonumber \\
     \dot{p}_\phi&=-\frac{2G\mu}{5c^5}\,\Ib^{(5)}_{\b1\b2}r^2
      ,\nonumber \\
     \dot{p}_r&=-\frac{G\M\mu}{r^2}+\frac{p_\phi^2}{\mu
     r^3}-\frac{3}{10}\frac{G\M\mu}{r^4}\left[\ka_n(a_1^2-a_3^2)+\ka_n'(a_1'^2-a_3'^2)\right] -\frac{2G\mu}{5c^5}\,\Ib^{(5)}_{\b1\b1}r   \ ,
\end{align}
while Eqs. (\ref{hameom21b}) remain unchanged.
In the quasi-static approximation it is the time varying orbital STF
mass quadrupole tensor alone, which contributes to
$\Ib^{(5)}_{\ba\bb}$. Explicitly, the components of
$\Ib^{(5)}_{\ba\bb}$ read
\begin{align}
\label{quadru}
       \Ib^{(5)}_{\b1\b1}&=-\frac{8G\M}{3}\frac{p_r}{r^4}
       \left[4\,\frac{G\M}{r}+3\,\frac{p_r^2}{\mu^2}
       +18\,\frac{p_\phi^2}{\mu^2r^2}\right]
       \ , \nonumber \\
       \Ib^{(5)}_{\b2\b2}&=\frac{2G\M}{3}\frac{p_r}{r^4}
       \left[8\,\frac{G\M}{r}+6\frac{p_r^2}{\mu^2}
       +81\,\frac{p_\phi^2}{\mu^2r^2}\right]
       \ ,\nonumber \\
       \Ib^{(5)}_{\b3\b3}&=\frac{2G\M}{3}\frac{p_r}{r^4}
       \left[8\,\frac{G\M}{r}+6\,\frac{p_r^2}{\mu^2}
       -9\,\frac{p_\phi^2}{\mu^2r^2}\right]
       \ , \nonumber \\
       \Ib^{(5)}_{\b1\b2}&=-\frac{4G\M}{r^5}\,p_\phi
       \left[2\,\frac{G\M}{r}+9\frac{p_r^2}{\mu^2}
       -6\,\frac{p_\phi^2}{\mu^2r^2}\right] \ .
\end{align}
For numerical calculations it is useful to apply the following scaling:
\begin{align}
\label{scaling}
     t=\frac{G\M}{c^3}\tau,\qquad
     p_r=\mu c\tp_r,\qquad
     &
     p_{a_i}=\mu c\tp_{a_i},\qquad
     p_\phi=\frac{G\M\mu}{c}\tp_\phi,\qquad
     p_\be=\frac{G\M\mu}{c}\tp_\be \nonumber \\
     &
     a_i=\frac{G\M}{c^2}\ta_i,\qquad
     r=\frac{G\M}{c^2}\tr.
\end{align}
Introducing the equilibrium radius $R_0$ of a nonrotating
(spherical) polytrope of mass $M$ and polytropic index $n$ the
terms containing $P_c/\rho_c$ can
be expressed as 
\begin{align}
\label{hilf}
     \frac{k_1M}{n}\frac{P_c}{\rho_c}=
     \frac{GM^2}{(5-n)R_0}\left(\frac{R_0}{R}\right)^{3/n}.
\end{align}
The full set of differential equations governing the dynamics of the
binary system is then given by
\begin{align}
\label{dynamic}
     \dot{\tp}_{a_1}&=\frac{5}{2\ka_nC_1}\frac{\tp_\be^2}{\ta_1^3}
     +\frac{\ka_n}{5}\frac{\ta_1}{\tr^3}
     +\frac{1}{5-n}\frac{C_1}{C_2}\left[\frac{2}{\ta_1\tR_0}\left(\frac{\tR_0}{\tR}\right)^{3/n}-3\,\frac{\ta_1A_1}{\tR^3}\right]
     +\frac{4\ka_n}{75}\frac{\ta_1}{C_2}\frac{\tp_r}{\tr^4}
     \left[\frac{8}{\tr}+6\tp_r^2-9\,\frac{\tp_\phi^2}{\tr^2}\right]
     \ , \nonumber \\
     \dot{\tp}_{a_1}'&=\frac{5}{2\ka_n'C_2}\frac{\tp_\be'^2}{\ta_1'^3}
     +\frac{\ka_n'}{5}\frac{\ta_1'}{\tr^3}
     +\frac{1}{5-n'}\frac{C_2}{C_1}\left[\frac{2}{\ta_1'\tR_0'}
     \left(\frac{\tR_0'}{\tR'}\right)^{3/n'}
     -3\,\frac{\ta_1'A_1'}{\tR'^3}\right]
     +\frac{4\ka_n'}{75}\frac{\ta_1'}{C_1}\frac{\tp_r}{\tr^4}
     \left[\frac{8}{\tr}+6\tp_r^2-9\,\frac{\tp_\phi^2}{\tr^2}\right]
     \ , \nonumber \\
     \dot{\tp}_{a_3}&=-\frac{\ka_n}{5}\frac{\ta_3}{\tr^3}
     +\frac{C_1}{C_2}\frac{1}{5-n}\left[\frac{1}{\ta_3\tR_0}
     \left(\frac{\tR_0}{\tR}\right)^{3/n}
     -\frac{3}{2}\frac{\ta_3A_3}{\tR^3}\right]
     -\frac{4\ka_n}{75}\frac{\ta_3}{C_2}\frac{\tp_r}{\tr^4}
     \left[\frac{8}{\tr}+6\tp_r^2-9\,\frac{\tp_\phi^2}{\tr^2}\right]
     \ , \nonumber \\
     \dot{\tp}_{a_3}'&=-\frac{\ka_n'}{5}\frac{\ta_3'}{\tr^3}
     +\frac{C_2}{C_1}\frac{1}{5-n'}
     \left[
     \frac{1}{\ta_3'\tR_0}\left(\frac{\tR_0'}{\tR'}\right)^{3/n'}
     -\frac{3}{2}\frac{\ta_3'A_3'}{\tR'^3}\right]
     -\frac{4\ka_n'}{75}\frac{\ta_3'}{C_1}\frac{\tp_r}{\tr^4}
     \left[\frac{8}{\tr}+6\tp_r^2-9\,\frac{\tp_\phi^2}{\tr^2}\right]
     \ ,\nonumber \\
     \dot{\tp}_\be&=\dot{\tp}_\be'=0 \nonumber \\
     \dot{\tp}_\phi&=\frac{8\nu}{5}\frac{\tp_\phi}{\tr^3}
     \left[\frac{2}{\tr}+9\tp_r^2-6\,\frac{\tp_\phi^2}{\tr^2}\right]
     \ , \nonumber \\
     \dot{\tp}_r&=-\frac{1}{\tr^2}+\frac{\tp_\phi^2}{\tr^3}
     -\frac{3}{10\tr^4}\left[\ka_n(\ta_1^2-\ta_3^2)
     +\ka_n'(\ta_1'^2-\ta_3'^2)\right]
     +\frac{16\nu}{15}\frac{\tp_r}{\tr^3}
     \left[\frac{4}{\tr}+3\tp_r^2+18\,\frac{\tp_\phi^2}{\tr^2}\right]
     \ , \nonumber \\
     \dot{\ta}_1&=\frac{5}{2\ka_nC_1}\,\tp_{a_1}
     \ , \nonumber \\
     \dot{\ta}_1'&=\frac{5}{2\ka_n'C_2}\,\tp_{a_1}'
     \ , \nonumber \\
     \dot{\ta}_3&=\frac{5}{\ka_nC_1}\,\tp_{a_3}
     \ ,  \nonumber \\
     \dot{\ta}_3'&=\frac{5}{\ka_n'C_2}\,\tp_{a_3}'
     \ , \nonumber \\
     \dot{\be}&=\frac{5}{2\ka_nC_1}\frac{\tp_\be}{\ta_1^2}
     \ ,\nonumber\\
     \dot{\be}'&=\frac{5}{2\ka_n'C_2}\frac{\tp_\be'}{\ta_1'^2}
     \ ,\nonumber \\
     \dot{\tr}&=\tp_r
     \ , \nonumber \\
     \dot{\phi}&=\frac{\tp_\phi}{\tr^2}
     \ ,
\end{align}
where $C_1=\frac{\M}{M'}=\frac{M}{\mu}$ and
$C_2=\frac{\M}{M}=\frac{M'}{\mu}$.
For the numerical calculations shown below we shall assume that
integration starts at periastron, i.e. the initial values for the
orbit are given by  
\begin{align*}
    \tr(0)=\tilde{d}(0)(1-\eps(0)),\qquad
    \phi(0)=0,\qquad
    \tp_r(0)=0,\qquad
    \tp_\phi(0)=\sqrt{\tilde{d}(0)(1-\eps(0)^2)}\quad,
\end{align*}
$\tilde{d}$ being the semi-major axis of the orbit.
The quasi-static approximation requires all velocities inside
the stars to be small. In particular, the mean radius $\tR$ varies only
slowly with time. Hence we impose $\dot{\tR}(0)\stackrel{!}{=}0$, i.e.
$2\dot{\ta}_1(0)\ta_1(0)\ta_3(0)+\ta_1(0)^2\dot{\ta}_3(0)=0$. This
yields immediately a  relation between the generalized momenta corresponding to
$a_1$ and $a_3$ respectively,
\begin{align*}
      \dot{\tp}_{a_3}(0)=-\frac{\ta_3(0)}{\ta_1(0)}\,\tp_{a_1}(0),\qquad
      \dot{\tp}_{a_3}'(0)&=-\frac{\ta_3'(0)}{\ta_1'(0)}\,\tp_{a_1}'(0).
\end{align*}
Given suitable (i.e. small) values of $\dot{\ta}_i(0),\
\dot{\ta}_i'(0)$ and $\dot{\be}(0)$ respectively $\dot{\be}'(0)$, we
still have to fix  the radii $R_0$ an $R_0'$, respectively. $R_0$
represents the equilibrium radius of a spherically symmetric
(i.e. non-rotating)  polytrope
of mass $M$ and polytropic index $n$. Assuming that the rotating star
is in equilibrium at $t=0$, we can apply the equilibrium relation
between $R$ and $R_0$, which has been derived in \cite{Lai:1993b}\footnote{Of
course this does not mean that the star is in equilibrium once the
integration has started.}:
\begin{align}
\label{equilib}
     R_0&=R(0)\left[
     \frac{3\arcsin e(0)}{e(0)}\,(1-e(0)^2)^{1/6}
     \left(1-\frac{1}{e(0)^2}+\frac{\sqrt{1-e(0)^2}}{e(0)\arcsin e(0)}\right)\right]^{n/(3-n)},
\end{align}
where $e(0)=\sqrt{1-(a_3(0)/a_1(0))^2}$ is the eccentricity of the ellipsoid.

\section{Gravitational waveforms}

In suitable coordinates the gravitational field, observed in an
asymptotically flat space, can be expressed as \cite{Thorne:1980ru}
\begin{align}
\label{rad0}
    h_{ij}^{rad}=\frac{G}{Dc^4}\sum_{l=2}^\infty\sum_{m=-l}^l
    \left[
    \left(\frac{1}{c}\right)^{l-2}\,
    ^{(l)}\mI^{lm}(t-\frac{D}{c})T_{ij}^{E2,lm}(\Theta,\Phi)+
    \left(\frac{1}{c}\right)^{l-1}\,^{(l)}\mathcal{S}^{lm}(t-\frac{D}{c})
    T_{ij}^{B2,lm}(\Theta,\Phi)\right],
\end{align}
where $D$ is the source-observer distance and the indices $i,j$ refer
to Cartesian coordinates in the asymptotic space.  $\mI^{lm}$ and
$\mathcal{S}^{lm}$ are the  spherical radiative mass and
current multipole moments, respectively, while $T_{ij}^{E2,lm}$
and $T_{ij}^{B2,lm}$
represent the so called pure-spin tensor harmonics of electric and
magnetic type, respectively. Finally, the upper index $l$ denotes the
number of time derivatives\footnote{Exploiting the transverse
traceless character of the gravitational radiation it is possible to
introduce two polarization vectors $\vec p$ and $\vec q$ in the plane
orthogonal to the direction of propagation. This leads to the more
familiar $h_+,h_\times$ notation,
$$
h_+=\frac{p_ip_j-q_iq_j}{2}h_{ij}^{rad},\qquad
h_\times=\frac{p_iq_j+p_jq_i}{2}h_{ij}^{rad},
$$
where $h_+$ and $h_\times$ represent the two polarization states.}.
In leading order
approximation only the $l=2$ terms contribute to the gravitational
field, which reflects the quadrupole character of the gravitational
radiation, i.e.
\begin{align}
\label{rad1}
     h_{ij}^{rad}=\frac{G}{Dc^4}\sum_{m=-2}^2
    \ddot{\mI}^{2m}(t-\frac{D}{c})T_{ij}^{E2,2m}(\Theta,\Phi).
\end{align}
However, one might chose to use the more familiar STF mass and current
multipole moments rather then calculating the spherical radiative ones
which appear in Eq. (\ref{rad1}). In
particular, the STF mass multipole moments are related to the
spherical radiative ones by
\begin{align}
\label{rad2}
      \mI^{lm}(t)=\frac{16\pi}{(2l+1)!!}\left[\frac{(l+1)(l+2)}{2l(l-1)}\right]^{1/2} \Ib_{A_l}Y^{\ast\ lm}_{A_l},
\end{align}
where $A_l=i_1\dots i_l$ is a multi-index, while the $Y^{lm}_{A_l}$ are
defined as
\begin{align*}
       Y^{lm}:=(-1)^m(2l-1)!!\sqrt{\frac{2l+1}{4\pi(l-m)!(l+m)!}}(\delta^1_{\langle i_1}+i\delta^2_{\langle i_1})\cdots(\delta^1_{i_m}+i\delta^2_{i_m})\delta^3_{i_{m+1}}\cdots\delta^3_{i_l\rangle}.
\end{align*}
The brackets $\langle\dots\rangle$ denote symmetrization.
Inserting this into Eq. (\ref{rad2})  the spherical radiative mass
quadrupole moments contributing to the leading order gravitational
waveform read
\begin{align}
\label{hrad1}
    \mI^{20}&=4\sqrt{\frac{3}{5\pi}}\Ib^{(IS)}_{33} , \\
\label{hrad2}    
    \mI^{21}&=-2\sqrt{\frac{8\pi}{5}}(\Ib^{(IS)}_{13}-i\Ib^{(IS)}_{23})
    , \\
\label{hrad3}    
    \mI^{22}&=\sqrt{\frac{8\pi}{5}}(\Ib^{(IS)}_{11}-\Ib^{(IS)}_{22}-2i\Ib^{(IS)}_{12}).
\end{align}
In our model only the  $\mI^{20}$ and  $\mI^{22}$-components are
present in the leading order gravitational wave field, since
$\mI^{21}$ vanishes due to the symmetry of the binary
system. To calculate the gravitational waveforms
explicitly we need to know the second time derivatives of the STF
mass quadrupole moment in the inertial frame. As before we shall
compute the contributions of the star's  quadrupole moments and the
orbital terms separately.

\subsubsection*{Orbital contribution}

Consider inertial center of mass system. In this system the
nonvanishing components of the orbital STF mass quadrupole tensor are
given by Eq. (\ref{orbit}). Taking the second time derivatives and
inserting the Newtonian equations of motion (\ref{lageom23}) and
(\ref{lageom24}) we end up with
\begin{align}
\label{orbital}
    \ddot{\Ib}^{(IS)}_{11}&=-2\mu r\dot{r}\dot{\phi}\sin 2\phi
       +\frac{\mu}{3}(1+3\cos
       2\phi)\left[\dot{r}^2-\frac{G\M}{r}\right]+\frac{\mu
       r^2}{3}(1-3\cos 2\phi)\dot{\phi}^2 \nonumber \\
       &\hspace{0.5cm} -\frac{G\M\mu}{10 r^3}(1+3\cos
       2\phi)\left[\ka_n(a_1^2-a_3^2)+\ka_n'(a_1'^2-a_3'^2)\right]
    \nonumber \\
    &=-2\frac{p_rp_\phi}{\mu r}\sin 2\phi +\frac{1}{3\mu}(1+3\cos
       2\phi)\left[p_r^2-\frac{G\M\mu^2}{r}\right] +\frac{p_\phi^2}{3\mu
       r^2}(1-3\cos 2\phi) \nonumber \\
       &\hspace{0.5cm}
       -\frac{G\M\mu}{10 r^3}(1+3\cos
       2\phi)\left[\ka_n(a_1^2-a_3^2)+\ka_n'(a_1'^2-a_3'^2)\right]
       ,  \\        
    \ddot{\Ib}^{(IS)}_{22}&=2\mu r\dot{r}\dot{\phi}\sin 2\phi
       +\frac{\mu}{3}(1-3\cos 2\phi)\left[\dot{r}^2-\frac{G\M}{r}\right]
       +\frac{\mu r^2}{3}(1+3\cos 2\phi)\dot{\phi}^2\nonumber \\
       &\hspace{0.5cm}
       -\frac{G\M\mu}{10r^3}(1-3\cos
       2\phi)\left[\ka_n(a_1^2-a_3^2)+\ka_n'(a_1'^2-a_3'^2)\right]
    \nonumber \\
    &=2\frac{p_rp_\phi}{\mu r}\sin 2\phi+\frac{1}{3\mu}(1-3\cos
       2\phi)\left[p_r^2-\frac{G\M\mu^2}{r}\right]+\frac{p_\phi^2}{3\mu
       r^2}(1+3\cos 2\phi)\nonumber \\
       &\hspace{0.5cm} -\frac{G\M\mu}{10 r^3}(1-3\cos
       2\phi)\left[\ka_n(a_1^2-a_3^2)+\ka_n'(a_1'^2-a_3'^2)\right]
       ,
\end{align}
\begin{align}
    \ddot{\Ib}^{(IS)}_{33}&=\frac{2\mu}{3}
       \left[\frac{G\M}{r}-r^2\dot{\phi}^2-\dot{r}^2\right]
       +\frac{G\M\mu}{5r^3}\left[\ka_n(a_1^2-a_3^2)
       +\ka_n'(a_1'^2-a_3'^2)\right] \nonumber \\
    &=\frac{2}{3\mu}\left[\frac{G\M\mu^2}{r}
       -\frac{p_\phi^2}{r^2}-p_r^2\right] +\frac{G\M\mu}{5r^3}
       \left[\ka_n(a_1^2-a_3^2)+\ka_n'(a_1'^2-a_3'^2)\right]
       , \\
    \ddot{\Ib}^{(IS)}_{12}&=2\mu r\dot{r}\dot{\phi}\cos 2\phi
       +\left[\dot{r}^2-\frac{G\M}{r}-r^2\dot{\phi}^2
       -\frac{3G\M}{10r^3}\bigl(\ka_n(a_1^2-a_3^2)
       +\ka_n'(a_1'^2-a_3'^2)\bigr)\right]\mu\sin 2\phi \nonumber \\
    &=2\frac{p_rp_\phi}{\mu r}\cos 2\phi
       +\left[\frac{p_r^2}{\mu}-\frac{G\M\mu}{r}-\frac{p_\phi^2}{\mu
    r^2}-\frac{3G\M\mu}{10r^3}\bigl(\ka_n(a_1^2-a_3^2)+\ka_n'(a_1'^2-a_3'^2)\bigr)\right]\sin 2\phi\quad .
\end{align}
Note that tidal coupling introduces  an additional contribution to
$\ddot{\Ib}^{(IS)}_{ij}$. Since this contribution is very small it can
be neglected for elliptical orbits and for $\ddot{\mc
I}^{22}$. However, it must be taken into account when considering 
$\ddot{\mc I}^{20}$ for circular orbits, where $\ddot{\mc I}^{20}$
would vanish identically for a point-particle system\footnote{Due to
tidal interaction effects the orbit is modified which in turn leads to
a small, but nonvanishing $\ddot{\mc I}^{20}$-component of the
gravitational radiation field.}.

\subsubsection*{Stellar contributions}

To calculate the contribution of, say, star $M$, let us assume  the
origin of the inertial system to coincide with the center of the
star. The coordinates $X_i$ in the inertial system
and coordinates $x_i$ in the body-fixed system are then related by the
$O(3)$-transformation given in Eq. (\ref{trafo}), and the elements of
the star's STF mass quadrupole tensor read
\begin{align*}
    \Ib^{(IS),star}_{ij}=T^\dag_{i\al}(\ga)T^\dag_{j\be}(\ga)\Ib^{star}_{\al\be}.
\end{align*}
$\Ib_{\al\be}^{star}$ refers to the body -fixed system, where the STF
mass quadrupole tensor takes a particularly simple form. For the most
general case when all semimajor axes of the star are different,
the transformation to the inertial system yields\footnote{From now on we
submit the index \textit{star}.}
\begin{align}
    \Ib^{(IS)}_{11}&=\frac{1}{2}(I_{11}-I_{22})\cos 2\ga
    +\frac{1}{6}(I_{11}+I_{22}-2I_{33})
    ,\nonumber \\
    \Ib^{(IS)}_{22}&=-\frac{1}{2}(I_{11}-I_{22})\cos
    2\ga+\frac{1}{6}(I_{11}+I_{22}-2I_{33})
    , \nonumber \\
    \Ib^{(IS)}_{33}&=-\frac{1}{3}(I_{11}+I_{22}-2I_{33})
    ,\nonumber \\
    \Ib^ {(IS)}_{12}&=\frac{1}{2}(I_{11}-I_{22})\sin 2\ga
    \quad.
\end{align}
It is easy to see that this reduces to  
$I^{(IS)}_{11}=I^{(IS)}_{22}=-\frac{1}{2}I^{(IS)}_{33}=\frac{1}{3}(I_{11}-I_{33})$
for $a_1\equiv a_2$. A straightforward calculation yields
\begin{align*}
   \ddot{\Ib}^{(IS)}_{11}&=\left[
     \frac{1}{2}(\ddot{I}_{11}-\ddot{I}_{22})-2(I_{11}-I_{22})\dot{\ga}^2
     \right]\cos 2\ga -
     \left[
     2(\dot{I}_{11}-\dot{I}_{22})\dot{\ga}+(I_{11}-I_{22})\ddot{\ga}
     \right] \sin 2\ga
     +\frac{1}{6}(\ddot{I}_{11}+\ddot{I}_{22}-2\ddot{I}_{33})
     , \\
   \ddot{\Ib}^{(IS)}_{22}&=-\left[
     \frac{1}{2}(\ddot{I}_{11}-\ddot{I}_{22})-2(I_{11}-I_{22})\dot{\ga}^2
     \right]\cos 2\ga +
     \left[
     2(\dot{I}_{11}-\dot{I}_{22})\dot{\ga}+(I_{11}-I_{22})\ddot{\ga}
     \right] \sin 2\ga
     +\frac{1}{6}(\ddot{I}_{11}+\ddot{I}_{22}-2\ddot{I}_{33})
     ,\\
   \ddot{\Ib}^{(IS)}_{33}&=-\frac{1}{3}(\ddot{I}_{11}+\ddot{I}_{22}
     -2\ddot{I}_{33})
     , \\
   \ddot{\Ib}^{(IS)}_{12}&=\left[
     \frac{1}{2}(\ddot{I}_{11}-\ddot{I}_{22})-2(I_{11}-I_{22})\dot{\ga}^2
     \right]\sin 2\ga
     +\left[
     2(\dot{I}_{11}-\dot{I}_{22})\dot{\ga}+(I_{11}-I_{22})\ddot{\ga}
     \right]\cos 2\ga
     \quad.
\end{align*}
In the quasi-static approximation the 2$nd$ time derivative of
the components of the stellar mass quadrupole tensor in the body fixed system vanishes
($I_{ij}\approx 0$) and  we are left with
\begin{align}
\label{star1}
     \ddot{\Ib}^{(IS)}_{11}&=-\ddot{\Ib}^{(IS)}_{22}
     =-2(I_{11}-I_{22})\om^2\cos 2\ga -
     \left[2(\dot{I}_{11}-\dot{I}_{22})\om
     +(I_{11}-I_{22})\dot{\om}\right]\sin 2\ga
     ,\nonumber \\
     \ddot{I}^{(IS)}_{12}&=-2(I_{11}-I_{22})\om^2\sin 2\ga
     +\left[2(\dot{I}_{11}-\dot{I}_{22})\om
     +(I_{11}-I_{22})\dot{\om}\right]\cos 2\ga , 
\end{align}
where $\Omega=\dot{\ga}$. We should
emphasize that Eqs. (\ref{star1}) imply that $\ddot{I}^{(IS)}_{ij}$
vanishes for axially symmetric polytropic stars, i.e. for $a_1\equiv a_2$.
In other words, in the quasi-static approximation there is no direct
contribution of the stars to the components of the gravitational
field. Nevertheless the internal degrees of freedom give rise to
modifications of the binary's gravitational wave forms due to tidally
driven modifications of the orbital motion and hence of $\ddot{\mc I}^{2m}$.

\subsubsection*{Gravitational waveforms of the binary system}

As it has been pointed out before, in the case of axially symmetric stellar
components of the binary only the orbital mass quadrupole tensor
contributes to the components of $h_{ij}$, the actual gravitational
waveforms being modified due to tidal coupling. Inserting
Eqs. (\ref{orbital}) into the expressions for $\mc I^{2m}$ given by
Eqs. (\ref{hrad1})-(\ref{hrad3}), the components of the leading order
gravitational wave field read
\begin{align}
\label{wave1}
    \ddot{\mc I}^{20}&=4\sqrt{\frac{3\pi}{5}}\mu c^2
        \left[\frac{2}{3}
        \left(\frac{1}{\tr}-\tp_r^2-\frac{\tp_\phi^2}{\tr^2}\right)
        +\frac{1}{5\tr^3}
        \left(\ka_n(\ta_1^2-\ta_3^2)+\ka_n'(\ta_1'^2-\ta_3'^2)\right)
        \right] ,\\
\label{wave2}         
    \Re(\ddot{\mc I}^{22})&=\sqrt{\frac{8\pi}{5}}\mu c^2
        \left[-4\frac{\tp_r\tp_\phi}{\tr}\sin 2\phi+
        \left\{
        2\left(\tp_r^2-\frac{1}{\tr}-\frac{\tp_\phi^2}{\tr^2}\right)
        -\frac{3}{5\tr^3}
        \left(\ka_n(\ta_1^2-\ta_3^2)+\ka_n'(\ta_1'^2-\ta_3'^2)\right)
        \right\}\cos 2\phi
        \right] ,\\
\label{wave3}
    \Im(\ddot{\mc I}^{22})&=\sqrt{\frac{8\pi}{5}}\mu c^2
        \left[-4\frac{\tp_r\tp_\phi}{\tr}\cos 2\phi-
        \left\{
        2\left(\tp_r^2-\frac{1}{\tr}-\frac{\tp_\phi^2}{\tr^2}\right)-
        \frac{3}{5\tr^3}
        \left(\ka_n(\ta_1^2-\ta_3^2)+\ka_n'(\ta_1'^2-\ta_3'^2)\right)
        \right\}\sin 2\phi
        \right] \quad.
\end{align}
Note the presence of an additional quadrupole-quadrupole coupling term
in above equations. This term can be easily neglected for the
22-component of the gravitational radiation, but it will significantly
modify the 20-component. This is due to the symmetry of the binary,
the spins of the stars being aligned perpendicular to the orbital plane.

\section{Numerical results and discussion}

\begin{figure}
\psfrag{t/T0}{$t/T_0$}
\psfrag{a1}{\hspace{-0.8cm}$a_1\quad [G\M/c^2]$}
\begin{center}
\includegraphics[width=8cm]{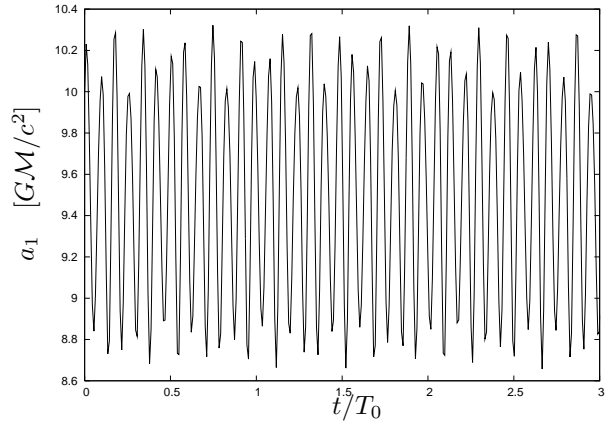}
\caption{Example for the stellar oscillation within the affine
model. Plotted is the oscillation of the semi-major axis $a_1$ of star
1 in a tidally coupled binary system.}
\label{fig:5}
\end{center}
\end{figure}

In the last few years there has been a growing interest in
investigating the influence of tidal interaction onto the inspiral
process and the actual leading order gravitational waveforms emitted
by a binary system. Within the framework of linear perturbation theory
of stellar oscillations, close binary systems of
nonrotating neutron stars where studied by Kokkotas and Sch\"afer
\cite{Kokkotas:1995xe} and later on by Lai and Ho \cite{Ho:1998hq}. In
these papers it was shown that tidal interaction may draw energy from
the orbital motion, thus speeding up the inspiral process. This effect
is strongest in the case of a tidal resonance when the orbital
frequency is the $m$th fraction of a star's eigenfrequency, $m$ being
an integer number. Resonant tidal oscillations of a nonrotating white
dwarf-compact object binary were investigated by Rathore et
al. \cite{Rathore:2004gs}, but in this paper dissipation due to
gravitational waves emission was not included.
On the contrary our
model incorporates not only dissipation due to gravitational wave
emission, but also stellar rotation. Moreover, we do not restrict
ourselves to small, linear perturbations of thermal equilibrium, but
allow for oscillations with arbitrary large amplitude. However, there is
only one oscillation mode per star present in our analysis (this mode
corresponding roughly to the f-mode), this means we truncate the
all the higher eigenmodes which are incorporated in the linear
perturbation scenarios. If much of the oscillation
energy is stored in these higher eigenmodes the gravitational
waveforms calculated from our model will fail to give results of high
accuracy. Therefore we restrict ourselves to moderate orbital
separations, when the dominant contribution of the oscillation energy
comes from the $f$-mode. In Fig. \ref{fig:5} the oscillation of the
semi-major axis $a_1$ is shown for a particular example.\\
The most prominent feature of the model is that it allows us to study
the long term evolution of the binary system. In fact, in our
numerical calculations we were able to follow the orbital evolution
over hundreds of periods. In particular we studied the influence of
the equation of state on the binary's dynamics and thus onto the
leading order gravitational wave pattern emitted during the time
evolution. In the
past it was argued that the influence of the polytropic index will not
be reflected in the gravitational waveforms almost until the final
plunge down \cite{Lai:1996sv}. While this is certainly true if one considers
only a few orbits this suggestion has to be carefully checked for long
term evolutions. Given suitable initial conditions our model allows
for the investigation of the orbital evolution over hundreds of
periods. It is thus an ideal tool to study the influence of the
polytropic index on the dynamics as well as on the gravitational
waveform in great detail. Note that the equations of motion given in
Eqs. (\ref{dynamic}) do not apply to the limit of incompressible fluids ($n=0$)
and to the relativistic limit, which is represented by $\Gamma=4/3$,
i.e. $n=3$. For $n=0$ Eq. (\ref{hilf}) has to be replaced by
\begin{align*}
    \frac{5k_1}{n\ka_n}\frac{P_c}{\rho_c}=2\frac{P_c}{\rho_c}.
\end{align*}
(For an explicit expression of $P_c/\rho_c$ in terms of the stellar
degrees of freedom see \cite{Lai:1995a}.) For $n\to 3$ numerical
integration becomes increasingly instable. Moreover,
Eq. (\ref{equilib}), which gives a relation between the equilibrium
radius $R_0$ of a nonrotating polytrope and the equilibrium mean
radius $R$ of a rotating polytrope, becomes singular for $n=3$. \\ 
In Fig. \ref{fig:1}  we compare the orbital evolution of a slightly elliptic
($\eps(0)=0.4$) equal mass binary system for different choices of
$n'$. As expected tidal coupling induces a periastron advance, but it is
clearly visible that the periastron advance increases with decreasing
values of $n'$. Not too surprisingly, the polytropic index $n'$ also
affects the inspiral process. For fixed value of $n$ the inspiral process
speeds up the smaller the values of $n'$, i.e. the larger the value of
$\Gamma'$.  
As it is shown in Figs. \ref{fig:2} and \ref{fig:4} the effect
of the equation of state on
the gravitational waveforms emitted by the binary \textit{cannot} be
neglected even for moderate orbital distances if one considers the
long-term evolution of the binary system. For circular orbits, the tidally
induced modification of the gravitational wave pattern is strongest
in the $\ddot{\mc I}^{20}-$component of $h_{ij}$ (see
Fig. \ref{fig:3}), but also in
the $22-$component there is an significant phase shift due to
different polytropic indizes $n'$. This
is demonstrated in Fig. \ref{fig:2} where we compare the
$\Re(\ddot{\mc I}^{22})-$component of the
gravitaional wave field for different values of $n'$. In these
particular examples the orbit is assumed to be circular in the absence
of tidal perturbation and the initial orbital separation is taken to
be $\tilde{d}(0)=50$. Even in this case a phase shift in the
$\Re(\ddot{\mc I}^{22})-$component is already obvious after 10 orbital
periods. \\
As already mentioned before, for a circular orbit the influence of the
equation of state is strongest reflected by the
the $\ddot{\mc I}^{20}$-component of $h_{ij}$.  For a point
particle binary, where the leading order gravitational waveform is
known analytically, this contribution to $h_{ij}$ would vanish for a
circular orbit. This
is not the case if the orbital motion is tidally coupled to the
internal motion of the stars (see Fig. \ref{fig:3}). \\
For elliptic binaries the influence of the polytropic index $n'$
on the $\ddot{\mc I}^{22}$-component is demonstrated in Figs.
\ref{fig:4} and \ref{fig:6}. Note that even  for large orbital
distances different polytropic indices $n'$ manifest themselves in a
remarkable phase shift of the $\ddot{\mc I}^{22}$-component of
$h_{ij}$ already after a few orbital cycles. This is demonstrated in
Fig. \ref{fig:6}, where we compare $\Re(\ddot{\mc I}^{22})$ for
different values of $n'$ at different stages of the inspiral
process.\\
To summarize, we have shown that including the internal structure
of the stars will lead to significant changes of the leading order
gravitational wave pattern compared to a point particle
binary. Moreover, even for moderate relative distances of the stars
there is a notable phase shift in the $\ddot{\mc I}^{22}$-component of
the gravitational radiation field after some orbital periods. Of
course our model does not respect all features of stellar dynamics. In
particular it does not account for higher order oscillation modes
which are present in linear perturbation theory. Nevertheless, our
model could give an approximative description of a  neutron
star-neutron star binary. Although the internal structure of a neutron star is
still not understood very good (the best known equations of motion
are only given in a tabulated form), astrophysical observations
indicate that the thermodynamics
of neutron stars can be approximately described by a polytropic
equation of state with polytropic index $n\approx 0.5-1.0$
(see e.g. \cite{Shapiro:1983}). Moreover, all low-mass white dwarfs
can be modelled with an effective polytropic index $n\approx
1.5$. Thus our model might be applicable to study the long term evolution
of NS-NS binaries, NS-White
Dwarf binaries or binaries consisting of either a neutron star or a
White Dwarf and a compact object.

\subsection*{Acknowledgements}

I am grateful to G. Sch\"afer and A. Gopakumar for many helpful
discussions and encouragement. This work was supported by the Deutsche
Forschungsgemeinschaft (DFG) through SFB/TR7
Gravitationswellenastronomie.

\begin{appendix}
\section{The Lane-Emden equation reviewed}

For spherically symmetric polytropes des equations of hydrodynamics
are given by
\begin{align*}
     \frac{dP}{dr}=-\frac{Gm(r)\rho(r)}{r^2},\qquad
     \frac{dm}{dr}=4\pi\rho(r)r^2
\end{align*}
which can be combined to a single equation
\begin{align}
\label{app1}
     \frac{1}{r^2}\frac{d}{dr}\left(\frac{r^2}{\rho}\frac{dP}{dr}\right)
     =-4\pi G\rho.
\end{align}
Introducing dimensionless variables $\xi$ and $\theta$ as
\begin{align*}
     \rho:=\rho_c\theta^n,\qquad r=a_0\xi,
\end{align*}
with
\begin{align*}
      a_0=\left[\frac{(n+1)K\rho_c^{\frac{1}{n}-1}}{4\pi G}\right]^{1/2},
\end{align*}
we obtain the  well known Lane-Emden equation
\begin{align}
\label{lame}
     \frac{1}{\xi^2}\frac{d}{d\xi}\left(\xi^2\frac{d\theta}{d\xi}\right)=-\theta^n,\qquad
     \theta(0)=1,\qquad \theta'(0)=0.
\end{align}
For $n<5$ the solution of Eq. (\ref{lame}) decreases monotonically and
becomes zero at a finite value of $\xi$. This value, denoted by
$\xi_1$, is characterized by vanishing pressure and density and hence
represents the star's surface. 

\section{Chandrasekhar's coefficients $A_i$ and $\mc J$}

The quantity $\mc J$ is defined in \cite{Chandrasekhar:1969} (chapter
3)  according
to
\begin{align}
\label{J}
     \mc J:=a_1a_2a_3\int_0^\infty \frac{du}{\sqrt{(a_1^2+u)(a_2^2+u)(a_3^2+u)}}
\end{align}
while the coefficients $A_i$ are given by
\begin{align}
\label{A}
     A_i:=a_1a_2a_3\int_0^\infty \frac{du}{(a_i^2+u)\sqrt{(a_1^2+u)(a_2^2+u)(a_3^2+u)}}.
\end{align}
One can easily show that
\begin{align*}
     \mc J=a_1^2A_1+a_2^2A_2+a_3^2A_3.
\end{align*}
Another useful relation that can be easily derived from Eq. (\ref{J}) is
\begin{align}
      \frac{\pa \mc J}{\pa a_i}=\frac{1}{a_i}(\mc J-a_i^2A_i)  \qquad
      (\text{no\ sum})
\end{align}
and
\begin{align}
\label{Jdiv}
      \frac{\pa \mc J}{\pa a_1}=\frac{2}{a_1}(\mc J-a_1^2A_1),\qquad
      \frac{\pa \mc J}{\pa a_3}=\frac{1}{a_3}(\mc J-a_3^2A_3)\qquad
      \text{for}\ a_1=a_2.
\end{align}
In the special case $a_1=a_2>a_3$ it is possible to find an analytical
expression for $A_1$ and $A_3$. Solving (\ref{A}) yields
\begin{align}
\label{Aspecial}
    A_1=A_2&=\frac{\sqrt{1-e^2}}{e^3}\,\arcsin \, e-\frac{1-e^2}{e^2}
    \\
    A_3&=\frac{2}{e^2}-\frac{2\sqrt{1-e^2}}{e^3}\,\arcsin\,e
\end{align}
where we defined
\begin{align}
     e:=\sqrt{1-\left(\frac{a_3}{a_1}\right)^2}.
\end{align}

\section{$\Ib^{(5),star}_{\ba\bb}$  beyond the quasi-static approximation}

Throughout the paper we applied the quasi-static approximation to
calculate the radiation reaction terms of the stellar degrees of
freedom. This approximation is justified if  all internal velocities
and accelerations inside the star are small. Dropping this assumption,
the result is much more
complicated. In particular, the 5$th$ time derivative of the STF mass
quadrupole tensor of $M$ in the corotating coordinate frame reads
\begin{align*}
     \Ib^{(5)}_{\ba\bb}&=
     (\Ib_{11}-\Ib_{22})
     \left[
     \bigl(16\om^5-40\om^2\ddot{\om}-60\om\dot{\om}^2+\om^{(4)}\bigr)
     \left(\begin{array} {ccc}
     \sin 2\al & \cos 2\al & 0 \\
     \cos 2\al & -\sin 2\al & 0 \\
     0 & 0 & 0 \end{array}\right)          \right.    \\
     &\hspace{2cm} - \left.
     \bigl(80\om^3\dot{\om}-20\dot{\om}\ddot{\om}-10\om\om^{(3)}\bigr)
     \left(\begin{array} {ccc}
      -\cos 2\al & \sin 2\al & 0 \\
      \sin 2\al & \cos 2\al & 0 \\
      0 & 0 & 0 \end{array}\right)
     \right] \\
     &+
     (\dot{\Ib}_{11}-\dot{\Ib}_{22})
     \left[
     \bigl(40\om^4-30\dot{\om}^2-40\om\ddot{\om}\bigr)
     \left(\begin{array} {ccc}
     \cos 2\al & \sin 2\al & 0 \\
     \sin 2\al & -\cos 2\al & 0 \\
     0 & 0 & 0 \end{array}\right)        \right. \\
     &\hspace{2cm} \left.+
     \bigl(120\om^2\dot{\om}-5\om^{(3)}\bigr)
     \left(\begin{array} {ccc}
     -\sin 2\al & \cos 2\al & 0 \\
     \cos 2\al & \sin 2\al & 0 \\
     0 & 0 & 0 \end{array}\right)
     \right] \\
     &+(\ddot{\Ib}_{11}-\ddot{\Ib}_{22})
     \left[
     (10\ddot{\om}-40\om^3)
     \left(\begin{array} {ccc}
     \sin 2\al & \cos 2\al & 0 \\
     \cos 2\al & -\sin 2\al & 0 \\
     0 & 0 & 0 \end{array}\right)
     +60\om\dot{\om}
     \left(\begin{array} {ccc}
     -\cos 2\al & \sin 2\al & 0 \\
     \sin 2\al & \cos 2\al & 0 \\
     0 & 0 & 0 \end{array}\right)
     \right] \\
     &+10(\Ib^{(3)}_{11}-\Ib^{(3)}_{22})
     \left[
     2\om^2\left(\begin{array} {ccc}
     -\cos 2\al & \sin 2\al & 0 \\
     \sin 2\al & \cos 2\al & 0 \\
     0 & 0 & 0 \end{array}\right)
     +\dot{\om}\left(\begin{array} {ccc}
     \sin 2\al & \cos 2\al & 0 \\
     \cos 2\al & -\sin 2\al & 0 \\
     0 & 0 & 0 \end{array}\right)
     \right] \\
     &
     +5\om(\Ib^{(4)}_{11}-\Ib^{(4)}_{22})
     \left(\begin{array} {ccc}
     \sin 2\al & \cos 2\al & 0 \\
     \cos 2\al & -\sin 2\al & 0 \\
     0 & 0 & 0 \end{array}\right) 
     +
     \frac{1}{2}(\Ib^{(5)}_{11}-\Ib^{(5)}_{22})
     \left(\begin{array} {ccc}
     \cos 2\al & -\sin 2\al & 0 \\
     -\sin 2\al & \cos 2\al & 0 \\
     0 & 0 & 0 
      \end{array}\right) \\
      &+
      \frac{1}{2}\left(\begin{array} {ccc}
      \Ib^{(5)}_{11}& & \\
      & \Ib^{(5)}_{22} & \\
      & & 2\Ib^{(5)}_{33} \end{array}\right).
\end{align*}

\end{appendix}

\begin{figure}[h]
\begin{center}
\begin{minipage}{0.45\linewidth}
\includegraphics[width=11cm]{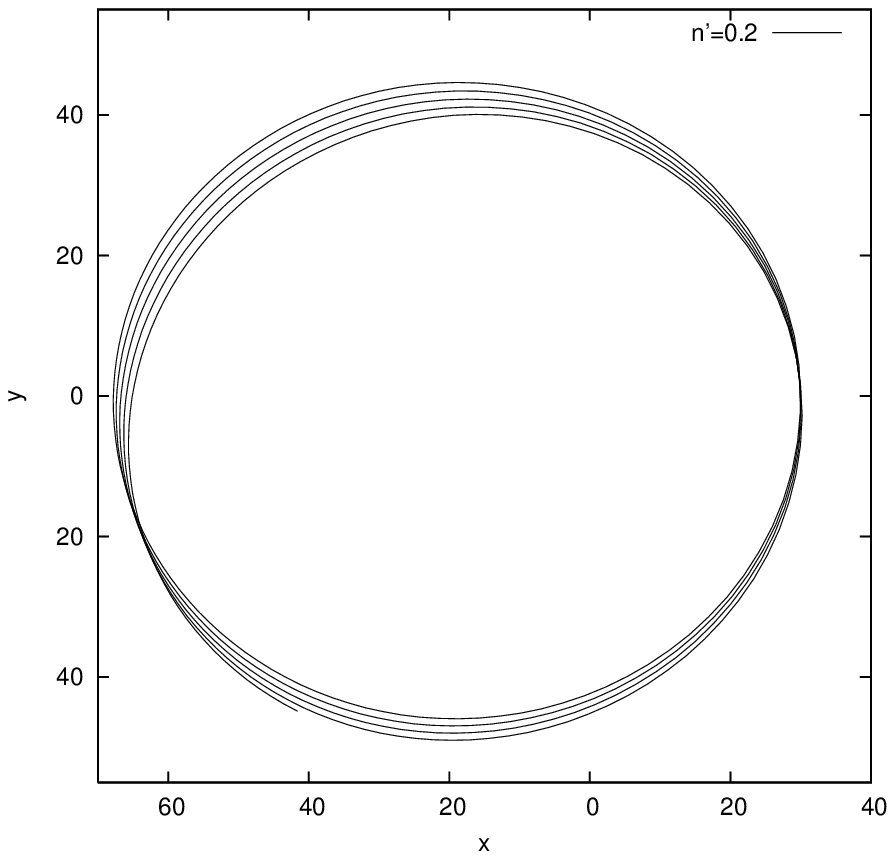}
\end{minipage}
\hfill
\begin{minipage}{0.45\linewidth}
\includegraphics[width=11cm]{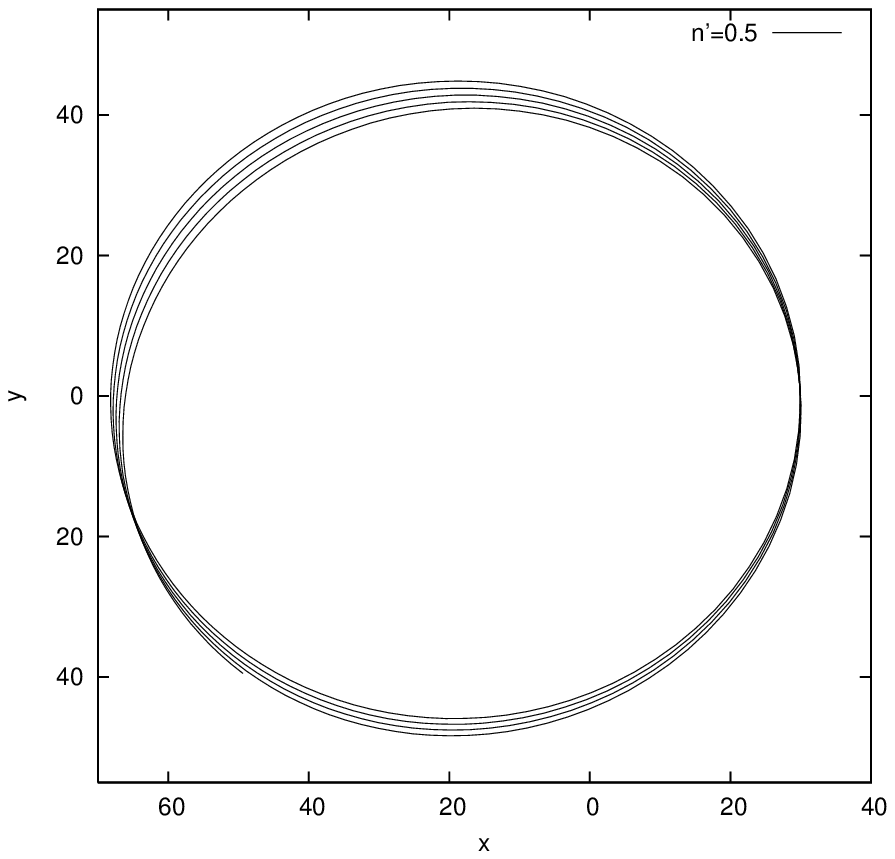}
\end{minipage}

\begin{minipage}{0.45\linewidth}
\includegraphics[width=11cm]{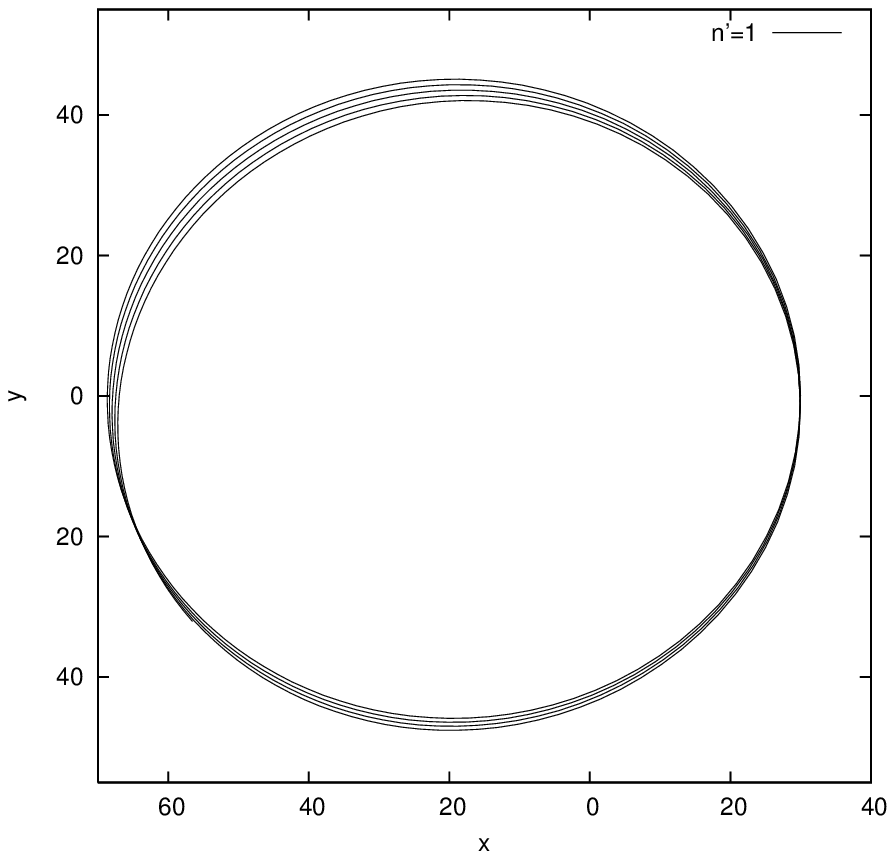}
\end{minipage}
\hfill
\begin{minipage}{0.45\linewidth}
\includegraphics[width=11cm]{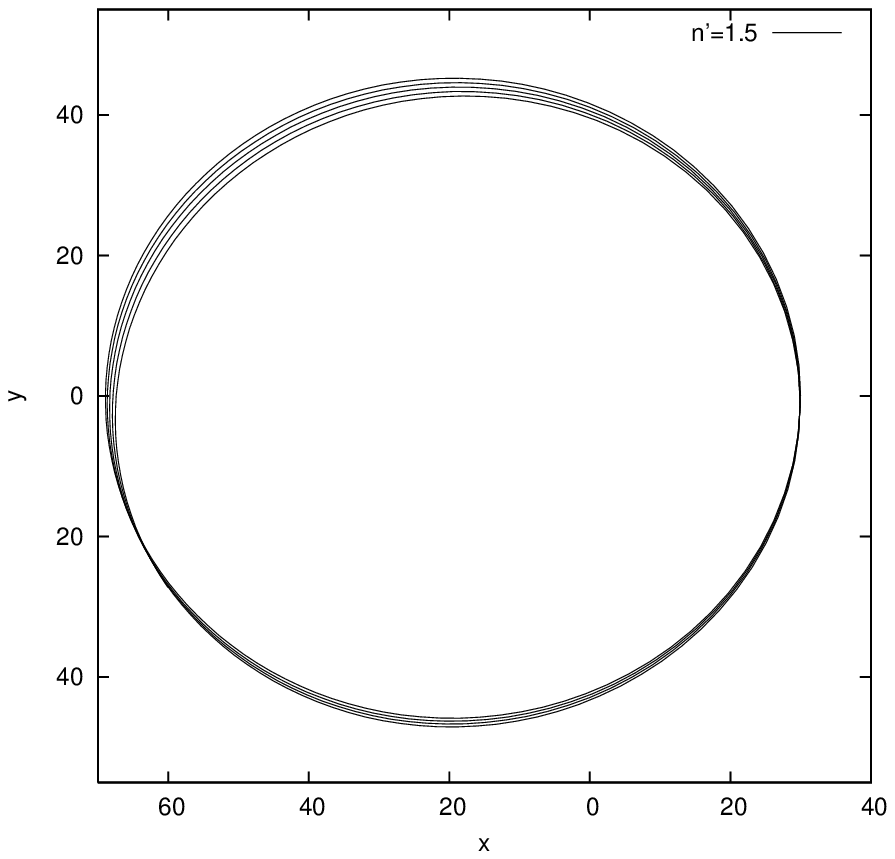}
\end{minipage}
\caption{Orbital evolution and periastron advance due to tidal
coupling for an elliptic, equal mass binary with initial values
$\eps(0)=0.4, \tilde{d}(0)=50, \ta_1(0)=\ta'_1(0)=10, \ta_3(0)=\ta'_3(0)=8,
\dot{\beta}(0)=\dot{\beta}'(0)=0.01, \dot{\ta}_1(0)=0.02,
\dot{\ta}'_1(0)=0.01$. All systems have polytropic index $n=1$, while
star 2 obeys a polytropic equation of state with $n'=0.2$ (upper
left), $n'=0.5$ (upper right), $n'=1$ (lower left) and $n'=1.5$ (lower
right). Star 1 has polytropic index $n=1$ in all scenarios.}
\label{fig:1}
\end{center}
\end{figure}

\begin{figure}[h]
\begin{center}
\psfrag{ReI22}{\hspace{-0.5cm}$\Re(\ddot{\mc I}^{22})\quad [\mu c^2]$}
\psfrag{t/T0}{$t/T_0$}
\hspace{-1cm}
\includegraphics[width=8.3cm]{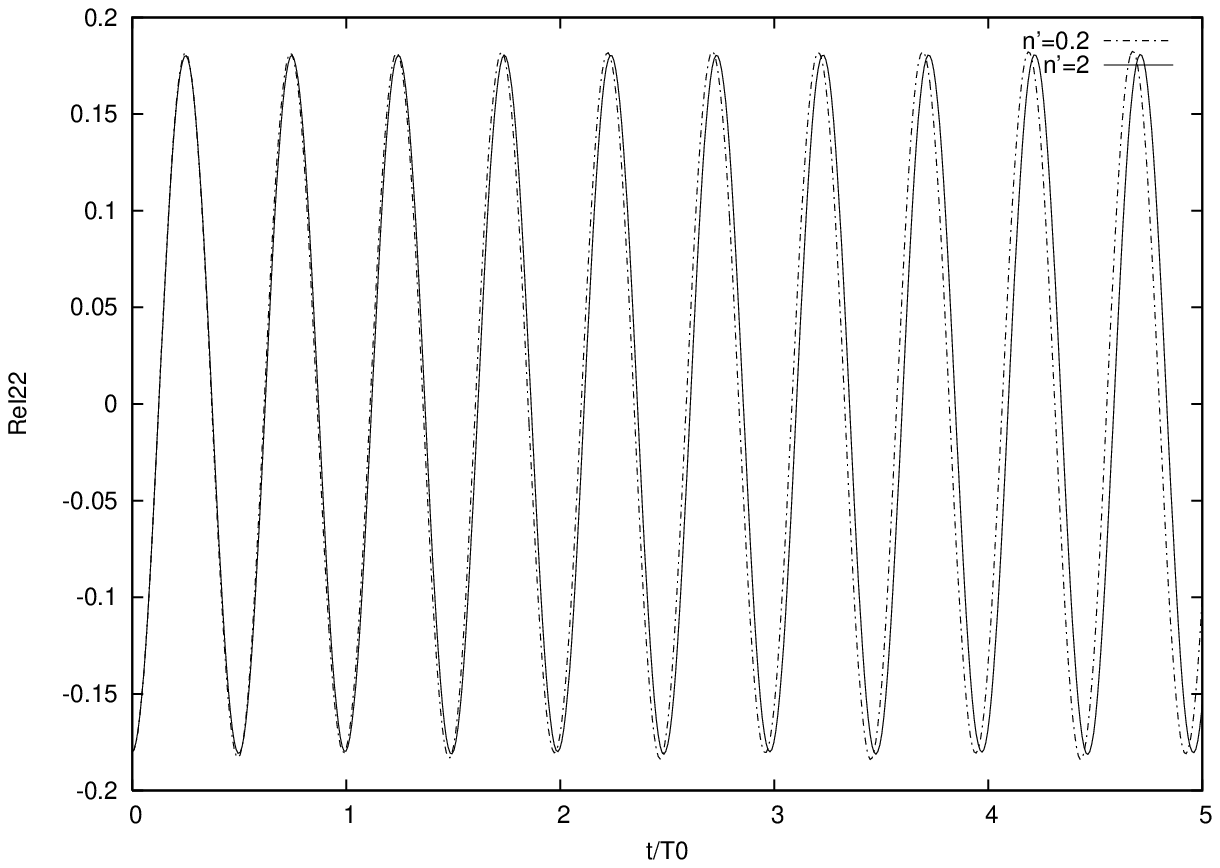}
\includegraphics[width=8.3cm]{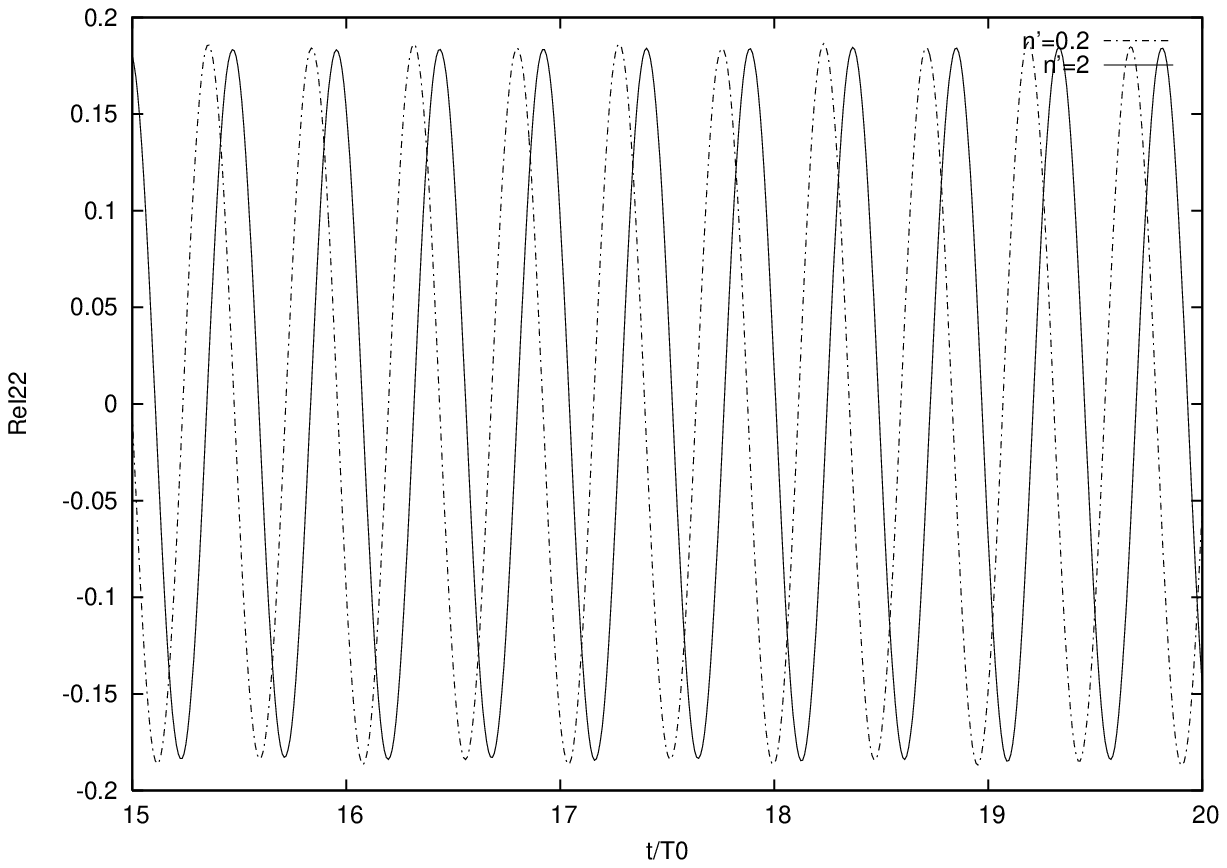}
\caption{The $\Re(\ddot{\mc I}^{22})$-component of the
leading order gravitational waveform for a circular orbit ($\eps(0)=0$) for
polytropic indices  $n'=0.2$ (dotted line) and
$n'=2$ (solid line) at different time stages. The polytropic index of
star 1 is assumed to be $n=1$,
the  initial parameters are the same as
in Fig. \ref{fig:1}.}
\label{fig:2}
\end{center}
\end{figure}

\begin{figure}
\begin{center}
\psfrag{I33}{\hspace{-0.5cm}$\ddot{\mc I}^{20}\quad [\mu c^2]$}
\psfrag{t/T0}{$t/T_0$}
\begin{minipage}{0.45\linewidth}
\hspace{-1.5cm}
\includegraphics[width=8.5cm]{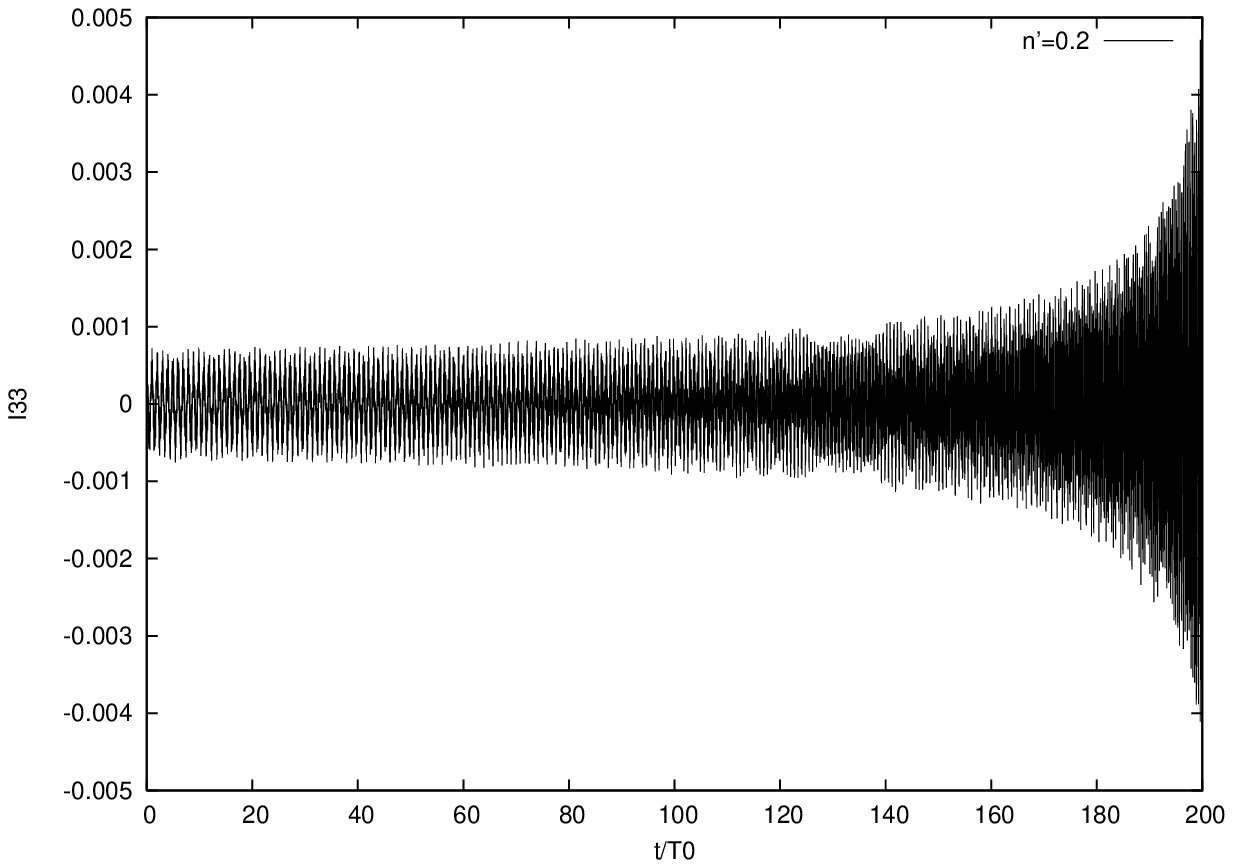}
\end{minipage}
\begin{minipage}{0.45\linewidth}
\includegraphics[width=8.5cm]{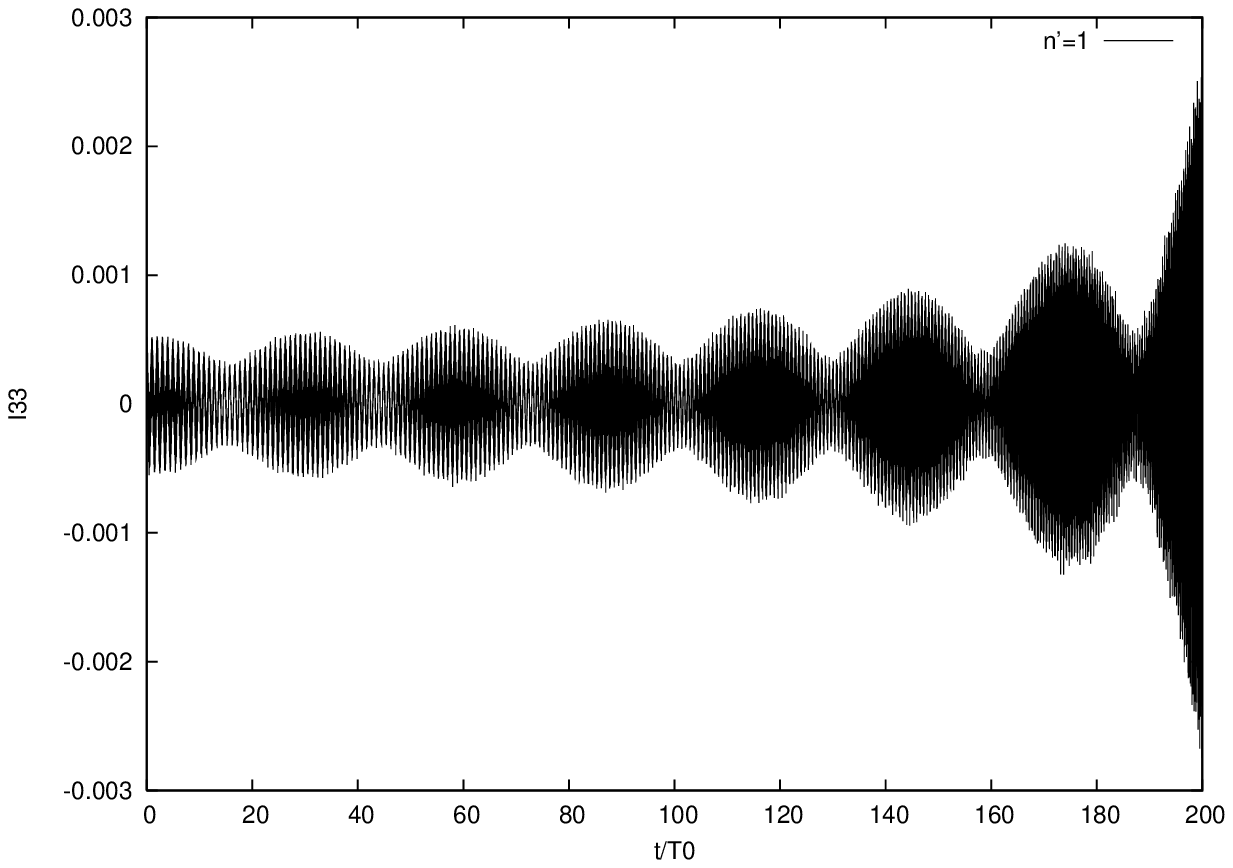}
\end{minipage}

\begin{minipage}{0.45\linewidth}
\hspace{-1.5cm}
\includegraphics[width=8.5cm]{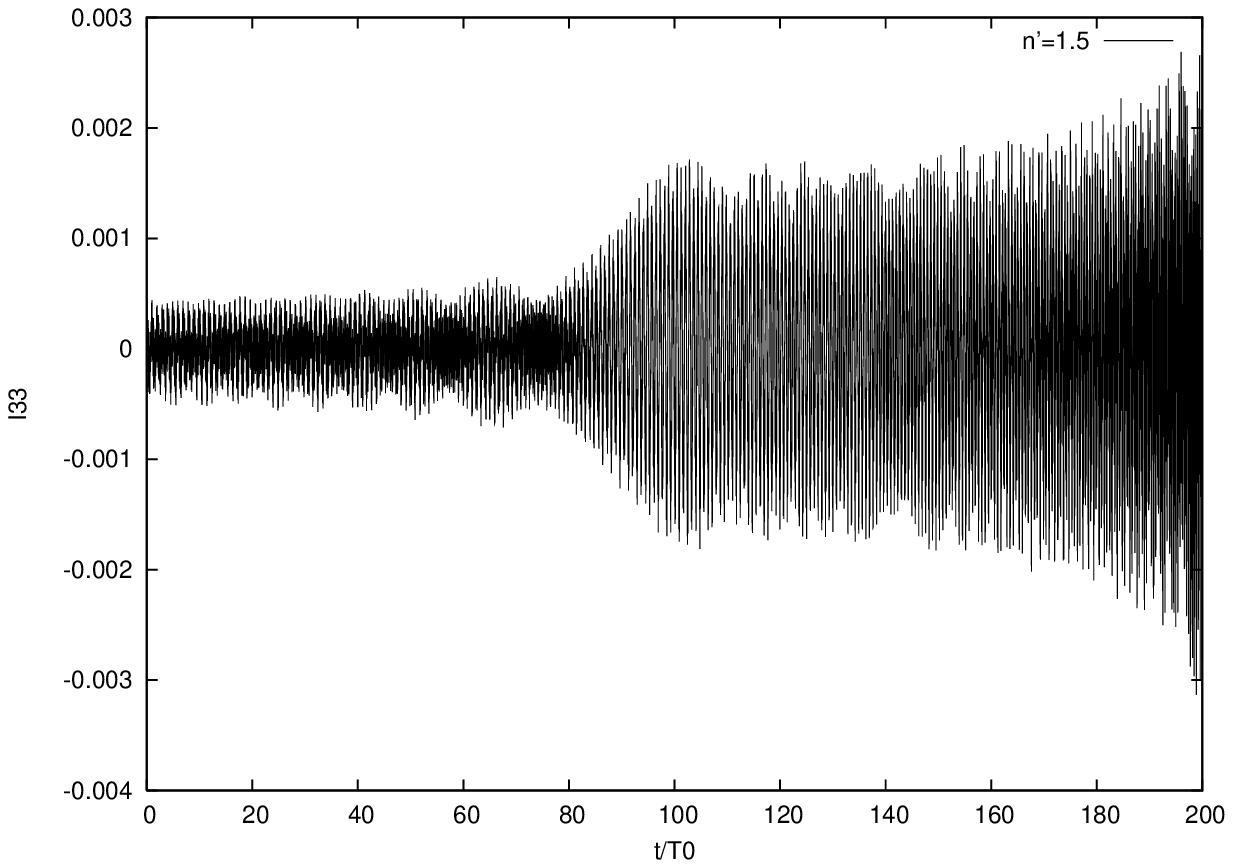}
\end{minipage}
\begin{minipage}{0.45\linewidth}
\includegraphics[width=8.5cm]{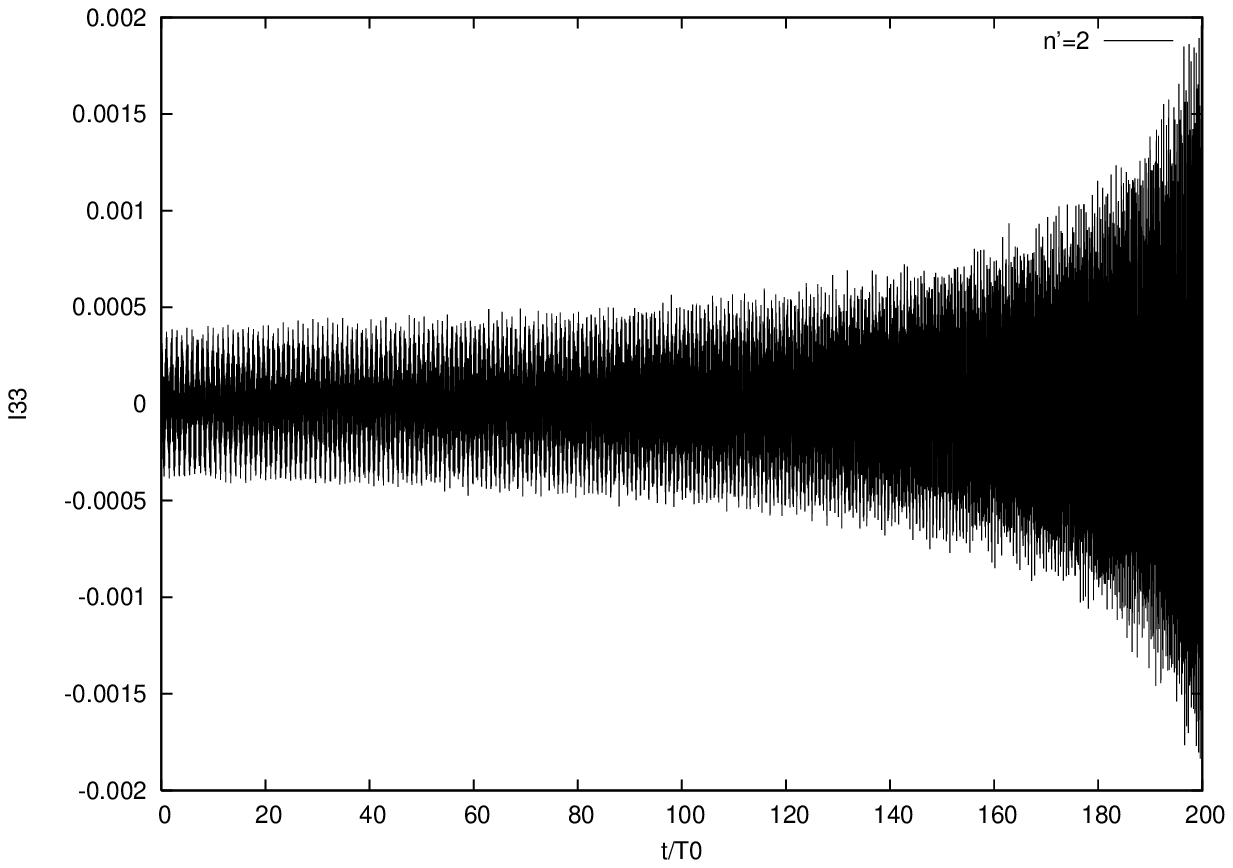}
\end{minipage}

\caption{$\ddot{\mc I}^{20}$-component of the leading order
gravitational waves for a binary with polytropic index $n=1$ for star
1 and
$n'=0.2$ (upper left), $n'=1$ (upper right), $n'=1.5$ (bottom left),
$n'=2$ (bottom right). The eccentricity at $t=0$ is assumed to be
zero, the initial parameters are the same as in Fig. \ref{fig:2}.}
\label{fig:3}
\end{center}
\end{figure}

\begin{figure}
\begin{center}
\psfrag{t/T0}{$t/T_0$}
\psfrag{ReI22}{\hspace{-0.5cm}$\Re(\ddot{\mc I}^{22})\quad [\mu c^2]$}
\hspace{-1cm}
\includegraphics[width=8.3cm]{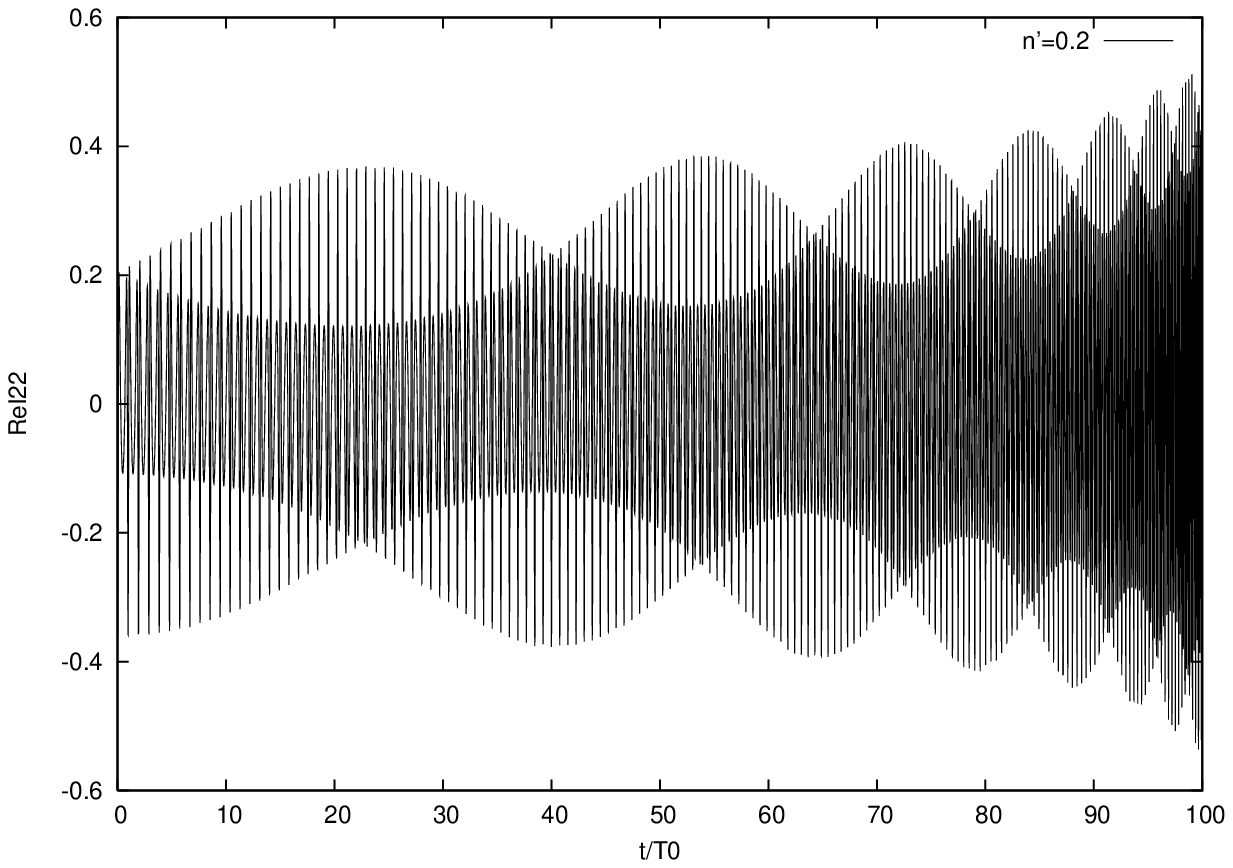}
\hfill
\includegraphics[width=8.3cm]{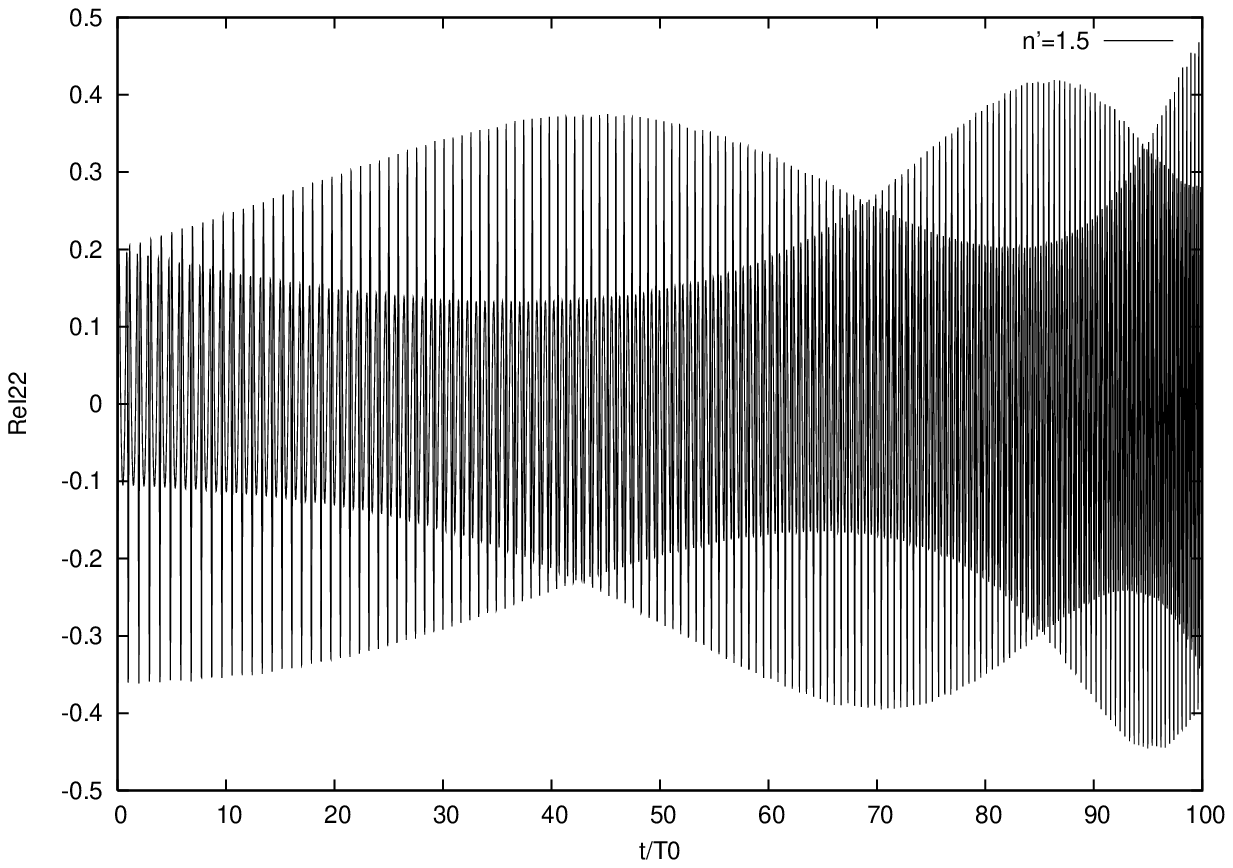}
\caption{Influence of the polytropic index $n'$ on the real part of
$\ddot{\mc I}^{22}$ for an elliptic binary with $\eps(0)=0.4$. The
initial parameters are the same as in Fig. \ref{fig:2}. Left: $n'=0.2$,
Right: $n'=1.5$.}
\label{fig:4}
\end{center}
\end{figure}

\begin{figure}
\begin{center}
\psfrag{t/T0}{$t/T_0$}
\psfrag{ReI22}{\hspace{-0.7cm}$\Re(\ddot{\mc I}^{22})\quad [\mu c^2]$}
\hspace{-1cm}
\includegraphics[width=8.2cm]{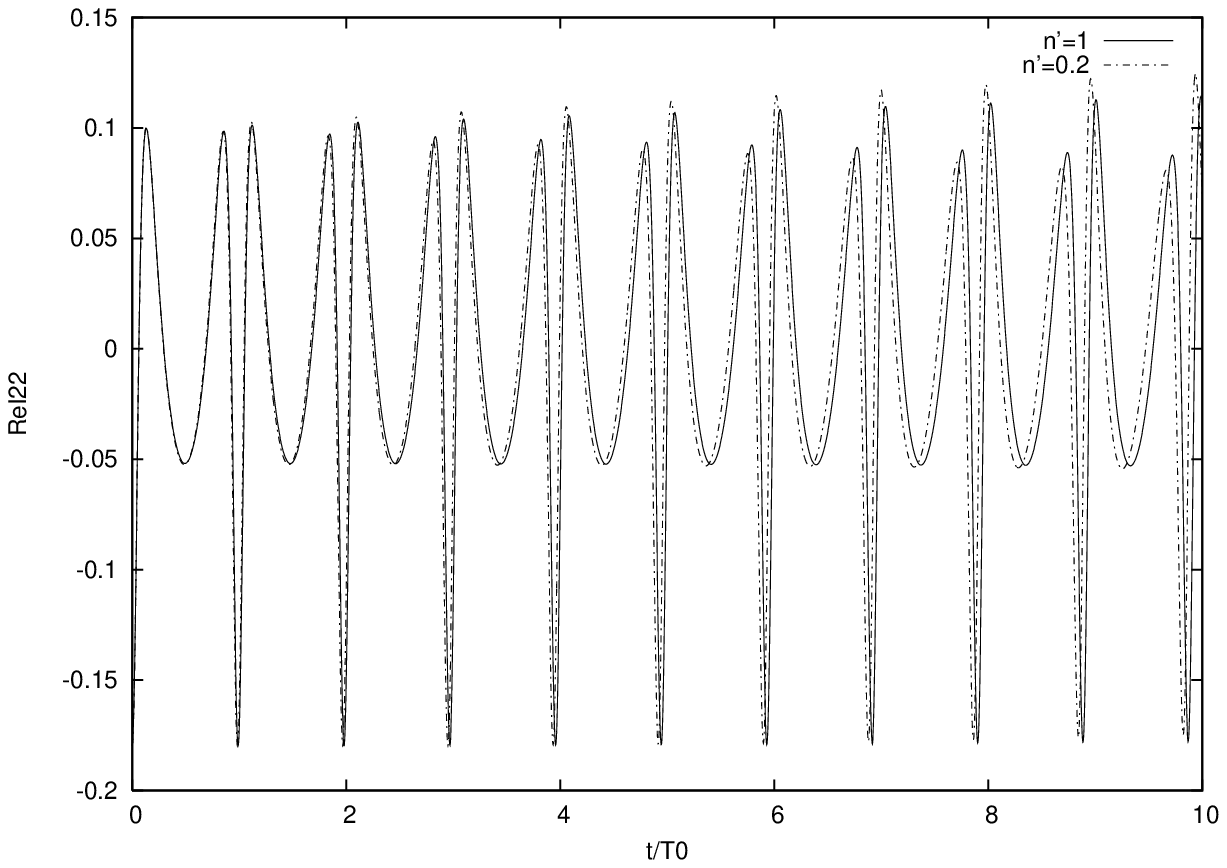}
\hfill
\includegraphics[width=8.2cm]{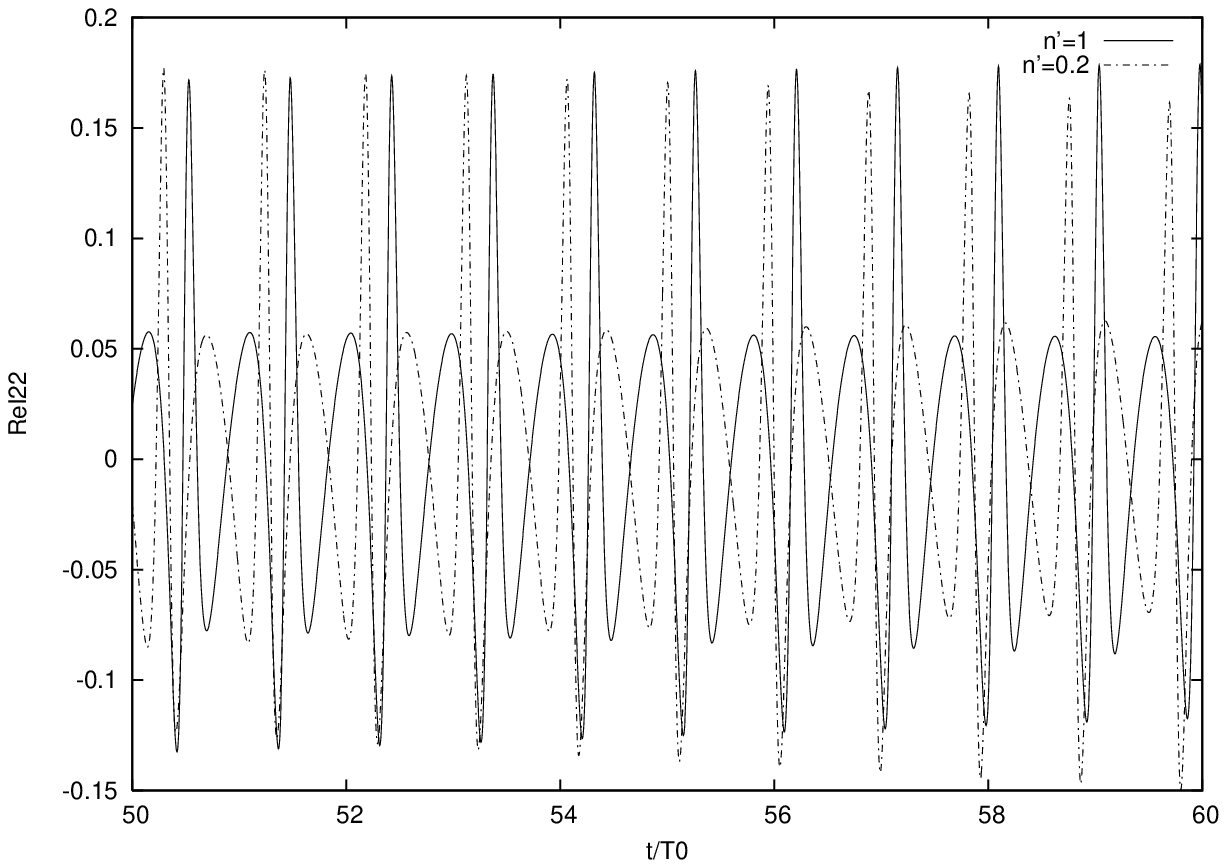}
\caption{$\Re(\ddot{\mc I}^{22})$-component of the leading order
gravitational radiation field emitted by an elliptic, equal-mass binary
($\eps(0)=0.4$). Shown are the waveforms for $n'=1$ (solid line) and
$n'=0.2$ (dotted line), the polytropic index of star 1 is assmued to
be $n=1$.  The semi-major axis at the beginning of the orbital
evolution  is taken to be $\tilde{d}(0)=100$, while the initial
parameters for the stars are given by $\ta_1(0)=\ta_1'(0)=20,
\ta_3(0)=\ta_3'(0)=19, \dot{\ta}_1(0)=\dot{\ta}'_1(0)=0$
and $\dot{\beta}(0)=0.002$, $\dot{\beta}'(0)=-0.002$. }
\label{fig:6}
\end{center}
\end{figure}

\end{document}